\documentclass[%
reprint,
 amsmath,amssymb,
 aps,
]{revtex4-2}

\usepackage[utf8]{inputenc} 
\usepackage{graphicx}
\usepackage{dcolumn}
\usepackage{bm}
\usepackage{xcolor}
\usepackage{miller}  
\usepackage{siunitx}
\usepackage{float}
\usepackage{subcaption}

\begin{document}

\preprint{APS/123-QED}

\title{Model for the Thermodynamics of Iron at High Pressures Near Melting}
\author{Rann Shikler}
 \email{rann.shikler@mail.huji.ac.il}
\author{Noya Dimanstein Firman}
\author{Yinon Ashkenazy}
 \email{yinon.ash@mail.huji.ac.il}
\affiliation{Racah Institute of Physics, Hebrew University of Jerusalem, Jerusalem 91904, Israel}

\date{\today}

\begin{abstract}
Iron phase diagram and specifically, its melting line at high pressures, have a wide range of applications, including the study of Earth’s inner core boundary (ICB). We propose an equation of state (EOS) model based on the interstitial theory of simple condensed matter\cite{granato92} (ITCM) focused on melting at high pressures. 
Melting is described as resulting from a strong structural perturbation due to interstitial-like defects. The nonlinearity of their interaction leads to the stabilization of interstitial-rich phase. The model is expanded to describe melting over a wide range of pressures.
The model is fitted to measured melting data and extrapolated to cover ICB conditions. The ITCM naturally yields two distinct liquid phases:  HCP and FCC-based liquids.  These phases and the transition between them are consistent with previous measurements. The additional liquid phase offers a new interpretation of the previously suggested near-melting high-pressure phase \cite{Boehler86}  and the inner core nucleation paradox \cite{HUGUET20189}.

\end{abstract}

\maketitle

\onecolumngrid

\section{Introduction} 
Iron's behavior at extreme pressures and temperatures is important in materials science and geophysics.
Iron, the most abundant element in Earth's core, has a melting line and thermodynamic properties essential to understanding processes such as planetary formation, magnetic field generation, and seismic activity \cite{Hirose2013, Stacey2005, Yasuhiro2020, Dorogokupets2017, zhang22, Aubert2013, LANDEAU2017}. 

Over the past decades, considerable effort has been devoted to constraining the melting temperature of Fe under Earth's core conditions.
Dynamic shock-loading experiments \cite{Brown&McQueen, Yoo&Holmes, Nguyen2004, Hermand, Balugani2024} and laser-heated diamond anvil cells (DAC) \cite{Anzellini2013, SINMYO2019, Boehler1993, Aquilanti2015, Tateno2010} have provided direct experimental 
access to extreme pressures and temperatures.
However, these methods often yield markedly different results, with discrepancies of up to 3000 K in the estimated melting temperature at the inner core boundary (ICB) \cite{Boehler1993}.
These variations largely reflect methodological limitations: laser-heated DAC experiments have limitations in temperature and pressure measurement \cite{Anzellini2020}, while shock experiments may probe metastable states due to kinetic effects \cite{Zeldovich, Duvall77}.

Melting of Iron at high pressures was also studied extensively using Density functional theory (DFT) calculations \cite{alfePRL, alfePRB}. 
The DFT results demonstrate a significant gap, and some of these results were incorporated into the fitting procedure of multiphase EOS models \cite{Livermore2023, Dorogokupets2017, saxena_2015}. 
Possible explanation for the divergence between experimental observations is based on the formation of a stable high-pressure BCC phase, which was suggested theoretically  \cite{Smirnov2023, Smirnov2025} and might be observed in recent experiments. \cite{konopkova2025}
However, the stabilization of such a phase and its possible role in non-pure systems remains unclear \cite{Li2024}.

In this manuscript, 
we pursue a complementary route to explore the properties of Iron at these conditions. 
We adopt a phenomenological EOS framework,  
fitting theory to previously available data, assuming there is no additional high-pressure phase.
Our work builds on a class of models \cite{granato92, wallace97, Lennard-Jones, Cotterill, Kleinert, deWith23, Poirier1986} that describe the free-energy difference between solid and liquid phases by treating the liquid as a highly perturbed solid and by explicitly estimating the free-energy variation associated with microscopic perturbations. 
Consequently, these methods may allow for improved extrapolation of phase lines into regions that are not directly measured.

A variety of perturbations have been considered in this context. 
Still, many suffer from inherent contradictions—for instance, dislocation-based models require too high energy per atomic length to be generated thermally \cite{Poirier1986}. 
In contrast, vacancy-based models yield entropy per volume that is too low \cite{deWith23}.
To overcome these limitations, we focus on Granato’s interstitial theory of condensed matter (ITCM) \cite{granato94, granato92, granato2014}, in which melting is described as a transition to a state with a high concentration of interstitial-like defects stabilized by the strong nonlinearity of their self-interactions. 
This theory has been shown to reproduce the kinetic and thermodynamic properties of simple liquid metals. Molecular dynamics simulations of solids with high interstitial concentrations reproduce dynamics similar to those of liquids \cite{Nordlund2005}, and crystallization speeds at large supercooling were shown to be consistent with interstitial-limited diffusion both theoretically \cite{Ashkenazy2007} and experimentally \cite{Chan2009}. 

\section{Free Energy Model}
We extend the original model \cite{granato2014, granato92, granato94}, which describes an isobaric phase transition, to a generalized form that covers a range of pressures and reconstructs the melting line over a broad thermodynamic regime. This extension enables the extrapolation of the theory to higher pressures than those where the melt line is reliably known.

Following Granato \cite{granato92}, we assume that the change in free energy caused by the addition of a single self-interstitial can be approximated by:
\begin{equation}
    \Delta G = H_{\text{F}}-TS_{\text{F}}\label{Eq:DeltaG} 
\end{equation}
Where $H_{\text{F}}$ and $S_{\text{F}}$ are the formation enthalpy and the formation entropy of interstitials, respectively.  

The interstitial formation enthalpy is assumed to include two types of linear elastic contributions
\begin{equation}
  \frac{dH_{\text{F}}}{dc}=\left(\alpha_{1}\mu(P,T,c)+\alpha_{2} B(P,T)\right)\cdot V(P,T,c)    
    \label{Eq:formationEnergy} 
\end{equation}
where $V$, $P$ and $T$ are the volume, pressure and temperature, respectively, $B$ and $\mu$ are the bulk and shear moduli and $\alpha_{1}$ and $\alpha_{2}$  are linear coefficients, it is assumed that $\alpha_1 \gg \alpha_2$.
This is combined with a model, that most changes in the free energy due to strain are caused by changes in the formation enthalpy, which leads to the following expression for the shear modulus:
$\mu(P,T,c)=\mu_{0}(P,T)\cdot \exp{(-\alpha_{1}\beta c)}$, 
Where $\beta$ is a dimensionless free parameter that depends on the crystal structure, the resulting softening of the shear modulus with the intersitial concentration is consistent with experimental observations. \cite{granato92}

The expression in Eq.~\ref{Eq:formationEnergy} depends on the volume, for which we assume an exponential dependence on the interstitial concentration: $V(P, T, c) = V_{0}(P, T) \cdot \exp{(\xi c)}$. Where $\xi$ is a free parameter quantifying the relationship between interstitial concentration and volume change. This approximation was verified to produce a consistent expression with the volume obtained from the derivative of the free energy. 

The entropy is assumed to include four major contributions \cite{granato92, granato94, granato2014}: frequency resonance modes, frequency local modes, frequency change due to volume change, and configuration entropy. Combined together, we get:

\begin{equation}
  S_{\text{F}} = Nk_{B}c \left[3\gamma \xi + \frac{4.35\cdot \nu}{1+a\alpha_{1}\beta c} -4.25 + \ln\left(\frac{b}{\alpha_{1}\beta c}\right) \right]
    \label{Eq:Entropyexp}
\end{equation}

Where $\gamma$ is the Grüneisen coefficient, and $a$, $b$, and $u$ are parameters that depend on the material\cite{granato92}.

Using Eq.~\ref{Eq:formationEnergy},~\ref{Eq:Entropyexp}, the change in free energy of a crystalline system due to the addition of interstitials to a concentration of $c$ can be written as:
\begin{eqnarray}
    \label{Eq:total}
    \Delta G =\\ 
    &&V_{0} \left[\frac{\alpha_{1} \mu_{0}}{\xi-\alpha_{1} \beta} (e^{(\xi-\alpha_{1} \beta)c}-1)+\frac{\alpha_{2} B}{\xi} (e^{\xi c}-1)\right] \nonumber \\
    && + Nk_{B}c \left[3\gamma \xi + \frac{4.35\cdot \nu}{1+a\alpha_{1}\beta c} -4.25 + \ln\left(\frac{b}{\alpha_{1}\beta c}\right) \right] \nonumber   
\end{eqnarray}
This estimates the free energy of the perturbed crystalline state due to the addition of self-interstitials at a concentration of $c$.

It is assumed that the highly perturbed crystalline state is in fact the liquid state, and so $G_{Liquid} = G_{Solid} + \Delta G$. This approach is beneficial as it utilizes well-established EOS data and models for the solid phases, ensuring higher reliability.

\section{Fitting} 
\subsection{Solid Phases}

Developing the EOS and phase diagram requires first fitting the solid free energy functions, which serve as the reference states for the liquid phases.
We use a classical and simple analytic form, similar to those shown to be reliable at high pressures and temperatures \cite{Livermore2023, Dorogokupets2017}.
Accordingly, the Helmholtz free energy for each solid phase is written as a combination of three terms:

\begin{equation}
  F(\rho, T) = F_{cold}(\rho) + F_{ion}(\rho, T) + F_{electron}(\rho, T)
\label{Eq:helmholtz}
\end{equation}
Where $F_{cold}$ represents the $T$-independent "cold energy" and $ F_{ion}, F_{electron}$, represent changes to the free energy from ionic and electronic excitations, respectively. Treating ionic and electronic excitations as independent is a manifestation of the Born–Oppenheimer approximation, as is often assumed in EOS studies \cite{Wallace, Bushman, Livermore2023, Bushman1983}.
 
To represent the cold term $F_{cold}$ for each crystalline phase, we used the Vinet analytic form \cite{Vinet_1989, Livermore2023}
\begin{equation}
  F_{cold}(\rho) = E_{0} + \frac{4B_{0}}{\rho_{0}(B_{0}'-1)^{2}}[1-(1+X)\exp{(-X)}]
\label{Eq:cold}
\end{equation}
in which $X=\frac{3}{2}\left(B_{0}'-1\right)[(\rho_{0}/\rho)^{1/3}-1]$
where $E_{0}, B_{0} \text{ and } B'_{0}$ are the energy, bulk modulus, and the pressure derivative of the bulk modulus at the density of $ \rho_{0}$.

For the ionic-excitation term of the free energy $F_{ion}$, we used the Debye model \cite{Debye1912, Livermore2023} 
\begin{eqnarray}
    F_{\text{ion}}(\rho, T) = && k_B T \left(\frac{9}{8} \frac{\theta(\rho)}{T} + \right.  \\
                            && \left. 3 \ln \left\{ 1 - \exp \left[ - \frac{\theta(\rho)}{T} \right] \right\}+ D_{3}\left[ \frac{\theta(\rho)}{T} \right] \right) \nonumber
    \label{Eq:ion}
\end{eqnarray}
where $\theta(\rho)$ is the Debye temperature fitted to a polynomial and $D_{3}$ is the third Debye function.

For the electronic-excitation term of the free energy $F_{electron}$, we used a model by Bushman\cite{Bushman}
\begin{equation}
     F_{elc}(\rho, T) = C_{V}T \ln \left[ 1 + 0.5\frac{\beta_{e}T}{C_{V}} \left(\frac{\rho_{0}}{\rho}\right)^{\gamma_{e}}\right]
\label{Eq:AGAV}
\end{equation}
where $C_{V}, \beta_{e} \text{ and } \gamma_{e}$ are the electronic heat capacity, electronic heat capacity coefficient, and the electronic Grüneisen coefficient, respectively.

We fitted the free energies for the $\alpha$ (BCC), $\gamma$ (FCC), and $\varepsilon$ (HCP) phases to various experimental data. This includes: 1-bar T-dependent thermal-expansion \cite{Abdullaev_2020, Assael2006} and heat capacity (entropy) \cite{NIST_JANAF} data for the $\alpha$ and $\gamma$ phases, Diamond Anvil Cell (DAC)\cite{Dewaele2008} and principal Hugoniot measurements \cite{Barker1974,Brown2000}, for $\alpha$ and $\varepsilon$ phases, heated Hugoniot \cite{Chen1997} for $\gamma$, ramp-compression data \cite{Smith2018} for $\varepsilon$ and the phase transition lines data \cite{Boettgaer1997, Swartzendruber1982}. The fitted parameters, fit results, and comparisons are in Appendix.~\ref{Appendi:Solid}

\subsection{Liquid Phases}
To model the liquid state EOS, we use the fitted multiphase solid EOS and the change in free energy due to the high concentrations of interstitials in Eq.~\ref {Eq:total}. 
Since the free energy of the liquid is described as a perturbed crystalline state, the fitting of the free energy was performed by identifying a solid state near melting and fitting a liquid phase transition using this phase as a reference.
The properties of the underlying crystalline phase are predetermined by the fit within its thermodynamic stability range.
This leaves as free-fitting parameters only those that describe interactions between the interstitials at high concentrations. 
We fitted liquid phases using only $\gamma$ (FCC) and $\varepsilon$ (HCP) melting lines, which dominate at high pressures and have been previously fitted. 
The BCC-related $\alpha$ and $\delta$ phases are stable at lower pressures and are excluded due to their limited relevance. 
The stable phase at each pressure and temperature was determined by identifying the phase with the minimum energy as a function of interstitial concentration for the explored solid phases.
This fitting leads to two separate liquid phases, namely, FCC and HCP-based liquids, denoted as $L_{FCC}$ and $L_{HCP}$.

The model was fitted to the melting line measurements from Sinmyo \cite{SINMYO2019}, as this dataset offers the broadest pressure range and approximates ICB conditions, enabling the most reliable extrapolation of the melting line. 
Because these measurements are the only ones above $200~(GPa)$, they dictate the high-pressure behavior of the fit when additional DAC datasets are included \cite{Anzellini2013,Boehler1993,Aquilanti2015}. In practice, adding these datasets produced only negligible changes, as the Sinmyo points lie within their overall spread.

To maintain thermodynamic consistency, Richard's rule \cite{richards_rule} was incorporated, governing the entropy change during melting. 
Additionally, bulk and shear moduli were independently fitted to a second-order pressure and first-order temperature expressions based on ab initio calculations \cite{Ai2020, coehn1997, soderlind1996}, allowing for extrapolation beyond the regime where measurements of the solid phases are available. 
We note that the validity of this EOS depends on both the extrapolation of the solid reference states' EOS and the model for the solid-liquid transition. 
Utilizing well-established analytical forms derived from physical consideration justifies the extrapolation of the free energies into the liquid regime. 
The transition into the liquid state addresses forming a perturbed state per solid phase.
However, a not well-defined upper limit on the validity of the model stems from the fact that the properties of the reference solid state are extrapolated to an untested regime and that we do not address the formation of new solid phases, such as a high BCC phase \cite{konopkova2025,Smirnov2023}

\begin{figure}[t!]
    \centering
    \includegraphics[width=12cm]{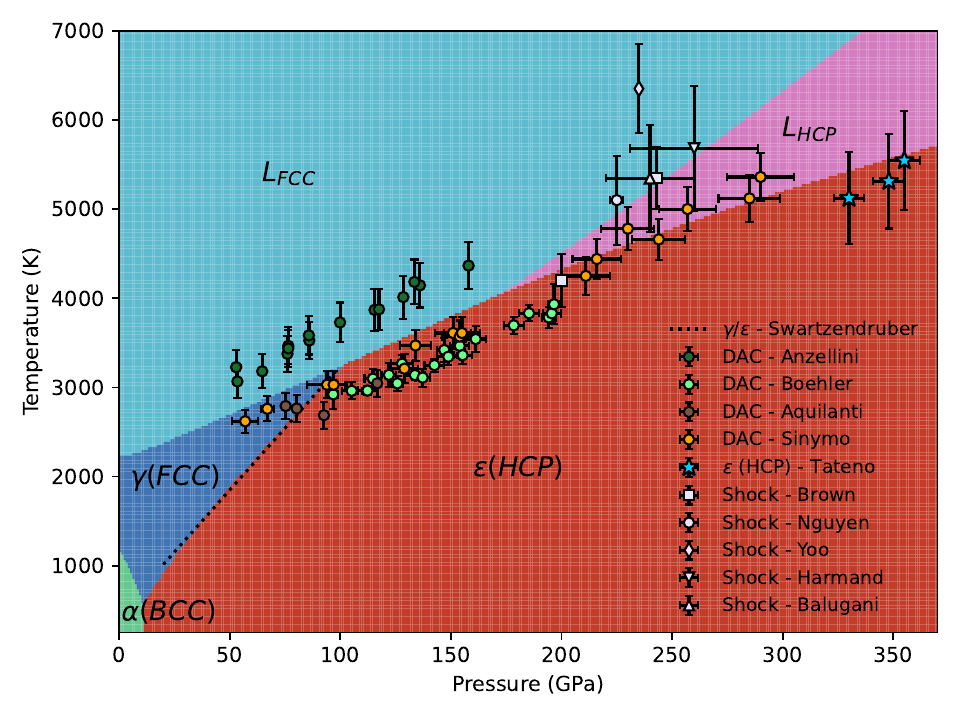}
    \caption{Fe phase diagram derived from our multiphase EOS. Light green indicates the $\alpha (BCC)$ phase, blue indicates $\gamma (FCC)$, red indicates $\varepsilon (HCP)$, light blue indicates $L_{FCC}$, pink indicates $L_{HCP}$. White markers indicate the Shock melt experiments, where each marker indicates a different experiment: square\cite{Brown&McQueen}, diamond\cite{Yoo&Holmes}, circle\cite{Nguyen2004}, downward triangle\cite{Hermand}, and upward triangle\cite{Balugani2024}. The black dashed line indicates the $\gamma-\varepsilon$ phase transition\cite{Swartzendruber1982}. The orange\cite{SINMYO2019}, green\cite{Anzellini2013}, light green\cite{Boehler1993} and gray\cite{Aquilanti2015} circle indicates the heated DAC melt experiments. The light blue star indicates heated DAC stability measurements of $\varepsilon(HCP)$\cite{Tateno2010}.}
    \label{fig:phases}
\end{figure}

\begin{figure}[t!]
    \centering
    \includegraphics[width=8.6cm]{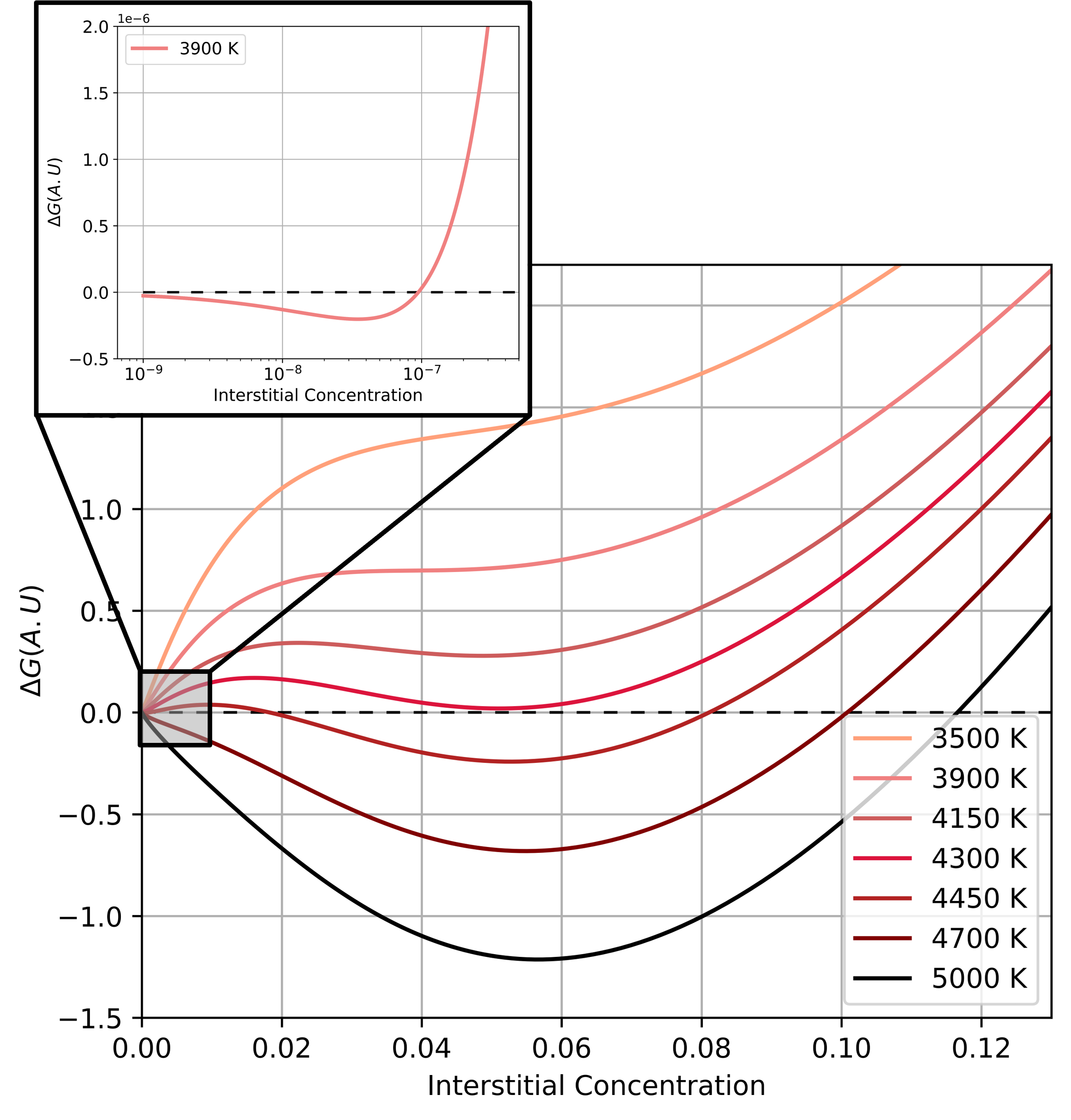}
    \caption{The change in Gibbs free energy as a function of the interstitial concentration  $\Delta G(c)$ at $200 [\text{GPa}]$ at different temperatures near melting.}
    \label{fig:interstitials}
\end{figure}

\section{Results and discussion} 

A comparison between the phase diagram derived from the fitted EOS and previous calculations and measurements of the Fe melting line is shown in Fig.~\ref{fig:phases}.
The melting line agrees with measurements by Sinmyo \cite{SINMYO2019}, while showing some deviation from earlier results by Anzellini, Boehler, and Aquilanti \cite{Anzellini2013, Boehler1993, Aquilanti2015}.
Overall, it lies near the center of the experimental spread.
Moreover, the phase boundaries are consistent with the high-pressure stability of $\varepsilon$-Fe (HCP) measurements reported by Tateno \cite{Tateno2010}.

Extrapolating the liquid phase free energies expression beyond their fitting range produced 
 a hypothetical $L_{FCC}-L_{HCP}$ phase transition between the two liquid phases. 
 The existence and location of this transition in the P-T phase diagram are not sensitive to a specific choice of fitting parameters. 
This phase transition agrees with results from shock melting experiments \cite{Brown&McQueen, Hermand, Nguyen2004, Balugani2024}, and may explain the discrepancy between melting measurements at high pressure. 
Specifically, it reproduces the explicit observation of two-phase transitions measured by Brown\& McQueen \cite{Brown&McQueen}. 
Such first-order liquid-liquid phase transitions were previously observed in experiments and atomistic simulations of other liquids, including liquid metals\cite{McMillan2000, Franzese2001}.

The fitted EOS generates free energy curves that maintain the properties exhibited in the original isobaric transition model\cite{granato92}.
Specifically, it leads to a stable solid phase with a low concentration of $\sim 10^ {- 7}$ up to the melting temperature. 
And a stable high interstitial concentration phase with about $6\%$ interstitial is expected to be amorphous and correspond to the liquid state, as illustrated in Fig.~\ref{fig:interstitials}.
At temperatures close to melting (within $\sim10\%$ of $T_M$), local minima in the free energy at high and low $c$ correspond to the existence of metastable liquid and solids.
At temperatures outside this region, $T < 0.9 T_M$ and  $T > 1.1 T_M$, $\Delta G_{s-l}(c)$ has only one minimum corresponding to the single stable phase.

\begin{figure}[t!]
    \centering
    \includegraphics[width=8.6cm]{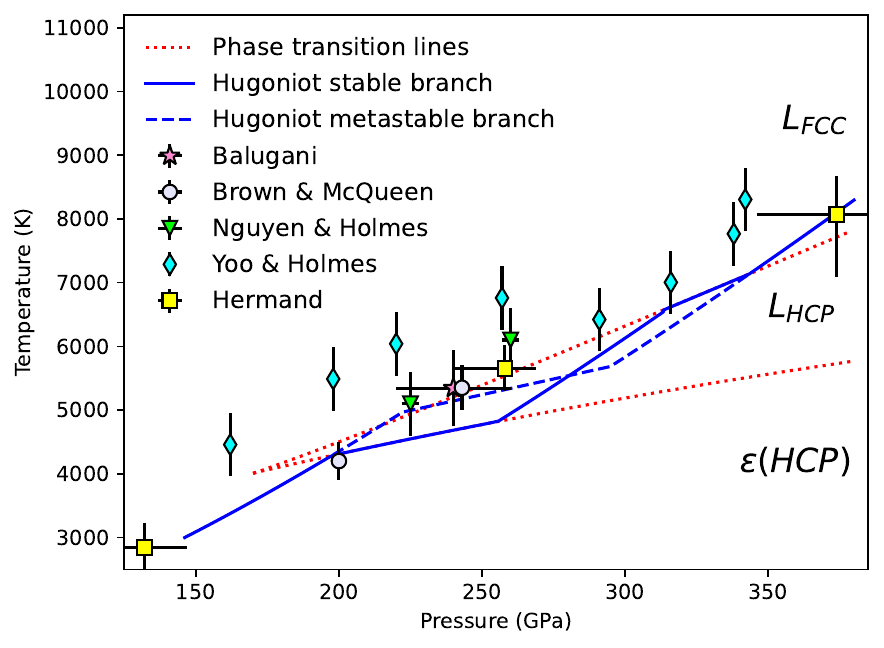}
    \caption{The Hugoniot possible trajectories near melting. The red dotted lines indicate the $\varepsilon - L_{HCP}$ and the $L_{HCP} - L_{FCC}$ phase transitions. The solid blue line represents the derived Hugoniot trajectory, while the dashed blue line indicates a metastable branch that passes through the edges of the solid stability regime. The markers represent different shock melting measurements: gray circles\cite{Brown&McQueen}, light blue diamonds\cite{Yoo&Holmes}, green triangles\cite{Nguyen2004}, yellow squares\cite{Hermand}, and pink stars\cite{Balugani2024}.}
    \label{fig:hugoniot}
\end{figure}

In Fig.~\ref{fig:hugoniot}, the Hugoniot curves derived from the model are compared with shock melt measurements.
The EOS leads to two closely spaced phase transition lines near melting at high pressures, one for melting and the other for the liquid-liquid phase transition.
These two phase transitions are expected to result in two separate transitions in the measured Hugoniot trajectory as it passes through the transition zone. Such a double transition was reported in shock compression measurement of elastic velocities  \cite{Brown&McQueen}.
The theoretical Hugoniot curve passing through these phases is shown in Fig.~\ref{fig:hugoniot}, which agrees with measurements below and above the melting and the liquid-liquid phase transition lines.
Between these lines, kinetics is expected to play a significant role as persistent metastable states are expected to be observed in dynamic experiments. 
The dashed line in Fig.~\ref{fig:hugoniot} represents the Hugoniot trajectory continuing in metastable solid state until the upper limit of solid metastability. This leads to an improved fit to the measured data at $250(GPa)$\cite{Brown&McQueen, Nguyen2004, Hermand, Balugani2024}.
The optional trajectories corresponding to different transition points may explain the spread in the experimental data.
Specifically, depending on kinetics, even a single transition from metastable solid to $L_{FCC}$, completely bypassing the intermediate liquid state, is possible.
The measurements by Yoo and Holmes \cite{Yoo&Holmes} deviate from our EOS, as shown in Figs.~\ref {fig:phases} and \ref{fig:hugoniot}. 
This discrepancy may be attributed to the model used in \cite{Yoo&Holmes} for the thermal capacity of the solid state, leading to differences in the estimated temperature, which is not directly measured.

Additional evaluations of melting at high pressures was done using Shock-Ramp  experiments at the National Ignition Facility (NIF) by Kraus et al. \cite{Kraus2022}. In these experiments, the phase at each measurement point was identified from the momentum transfer spectra, and deviations from classical liquid spectra were observed in a region near melting.
Using ab initio simulations of Iron it was suggested that the observed signal can be explained in terms of partial formation of a BCC phase at high temperature and pressure \cite{Smirnov2023}. 
 ITCM leads to an alternate hypothesizes,  
 where such deviations may result from increase in resonant and local vibrational modes which may be observed near melting. However this needs to be further explored using computational and experimental study.

In addition, seismic velocity measurements and simulations have shown that Earth’s inner core displays strong shear softening and an ultrahigh Poisson’s ratio.
In simulations of Fe HCP solid close to melting, this softening was attributed to the formation of increased correlation in the displacement of atoms, manifested in chains of collective motion.\cite{Zhang2023}
Such increased correlation leading to the formation of displacement chains was previously reported in atomistic simulations of Copper highly perturbed crystalline phase as melting was approached.\cite{Nordlund2005}
Thus, this observed softening may be related to the suggested intermediate $L_{HCP}$.
The formation of such an intermediate liquid phase is expected to reduce the nucleation barrier due to its effect on the solid-liquid interface energy\cite{Ashkenazy2007, Chan2009}. 
This, in turn, can explain the gap between the observed structure and the estimated nucleation time of the solid phase in Earth's inner core. \cite{HUGUET20189}.
Furthermore, the $L_{HCP}$ phase is a natural candidate as an intermediate phase for a two-step process, which was previously suggested to explain this phenomenon \cite{Two_step}.

\begin{figure}[H]
    \centering
    \includegraphics[width=8.6cm]{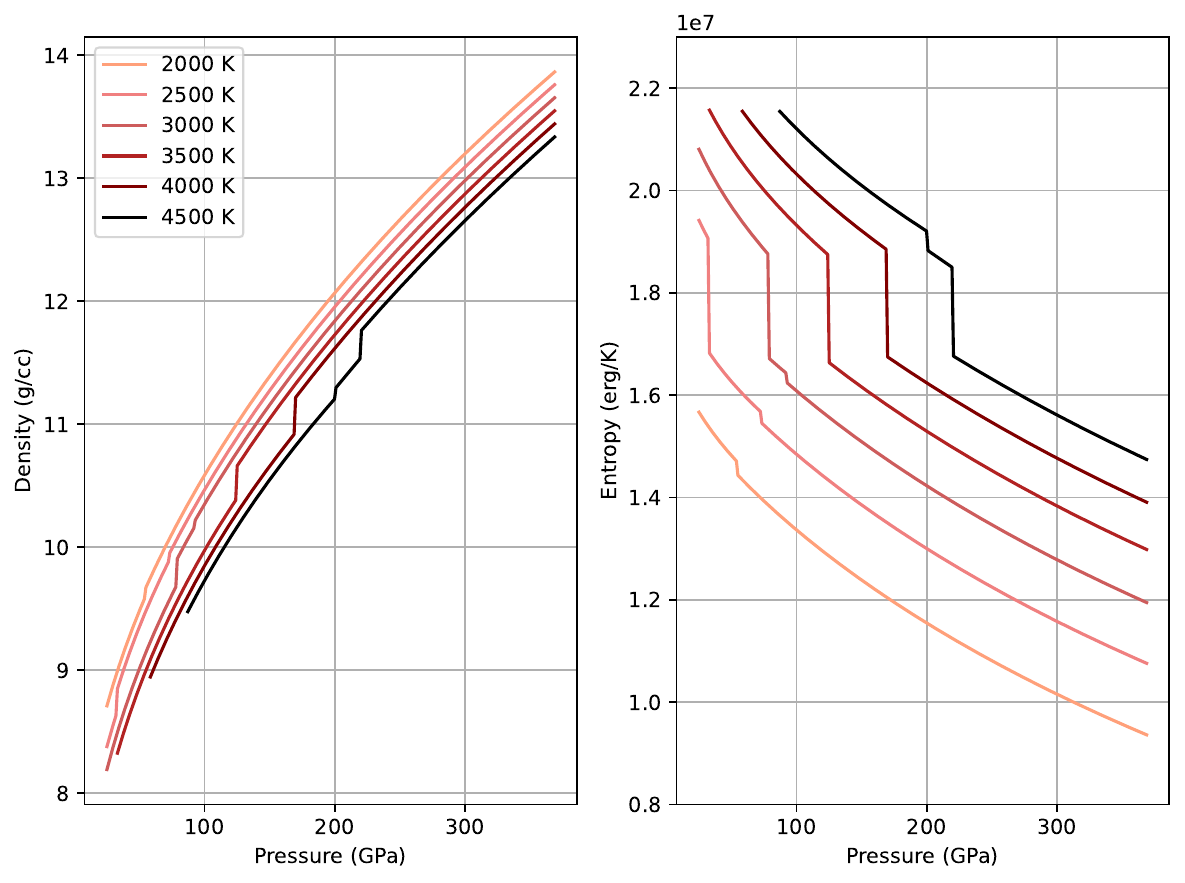}
    \caption{Isotherms of (a) the density $\rho$ and (b) the entropy $S$ per gram at different temperatures. The vertical lines represent phase transitions along the isotherms, where the larger jumps indicate melting.}
    \label{fig:Isotherms}
\end{figure}

Another type of thermodynamic trajectory that may be accessible experimentally is the compression isotherms of the EOS presented in Fig.~\ref{fig:Isotherms}. The model's entropy jump at melting is consistent with Richards' rule \cite{richards_rule}, and the density profile follows a reasonable trend. \cite{Dewaele2008}

\section{Conclusions} 
Using ITCM, we derived expressions for the free energy functions of various iron phases at high pressure and temperature. 
Applying ITCM with two different solid reference phases at low temperatures leads to a phase diagram that includes two liquid phases. Calculations based on these functions coincide with measurements and prior computations.

The predicted $L_{HCP}$-$L_{FCC}$ phase transition line at high pressures is consistent with the divergence of existing melting measurements at high pressures. Furthermore, the $L_{HCP}$ phase can explain the speculated high-pressure phase\cite{Boehler86, Brown&McQueen}.

We note that exploring the proposed liquid–liquid phase transition is expected to be challenging for computational techniques such as ab-initio calculations, as it requires reproducing intrinsic phenomena such as high entropy contributions from resonant modes.

ITCM's liquid descriptions can have significant implications for Earth studies, such as predicting the melting lines for Fe under ICB conditions in a different method, explaining Earth's inner core's strong shear softening and an ultrahigh Poisson’s ratio, and a possible solution to the Earth's inner core nucleation paradox.

This study represents the first application of ITCM with two reference states, achieving an improved fit to experimental data for iron under extreme conditions. The modified ITCM and derived free energy functions reveal a potential liquid-liquid phase transition, consistent with observations and offering new insights into iron's behavior at high pressures.


\section{Declaration of competing interest}
The authors declare that they have no known competing financial interests or personal relationships that could have appeared to influence the work reported in this manuscript.

\bibliography{references}

\providecommand{\noopsort}[1]{}\providecommand{\singleletter}[1]{#1}%
\begin{thebibliography}{67}%
\makeatletter
\providecommand \@ifxundefined [1]{%
 \@ifx{#1\undefined}
}%
\providecommand \@ifnum [1]{%
 \ifnum #1\expandafter \@firstoftwo
 \else \expandafter \@secondoftwo
 \fi
}%
\providecommand \@ifx [1]{%
 \ifx #1\expandafter \@firstoftwo
 \else \expandafter \@secondoftwo
 \fi
}%
\providecommand \natexlab [1]{#1}%
\providecommand \enquote  [1]{``#1''}%
\providecommand \bibnamefont  [1]{#1}%
\providecommand \bibfnamefont [1]{#1}%
\providecommand \citenamefont [1]{#1}%
\providecommand \href@noop [0]{\@secondoftwo}%
\providecommand \href [0]{\begingroup \@sanitize@url \@href}%
\providecommand \@href[1]{\@@startlink{#1}\@@href}%
\providecommand \@@href[1]{\endgroup#1\@@endlink}%
\providecommand \@sanitize@url [0]{\catcode `\\12\catcode `\$12\catcode `\&12\catcode `\#12\catcode `\^12\catcode `\_12\catcode `\%12\relax}%
\providecommand \@@startlink[1]{}%
\providecommand \@@endlink[0]{}%
\providecommand \url  [0]{\begingroup\@sanitize@url \@url }%
\providecommand \@url [1]{\endgroup\@href {#1}{\urlprefix }}%
\providecommand \urlprefix  [0]{URL }%
\providecommand \Eprint [0]{\href }%
\providecommand \doibase [0]{https://doi.org/}%
\providecommand \selectlanguage [0]{\@gobble}%
\providecommand \bibinfo  [0]{\@secondoftwo}%
\providecommand \bibfield  [0]{\@secondoftwo}%
\providecommand \translation [1]{[#1]}%
\providecommand \BibitemOpen [0]{}%
\providecommand \bibitemStop [0]{}%
\providecommand \bibitemNoStop [0]{.\EOS\space}%
\providecommand \EOS [0]{\spacefactor3000\relax}%
\providecommand \BibitemShut  [1]{\csname bibitem#1\endcsname}%
\let\auto@bib@innerbib\@empty
\bibitem [{\citenamefont {Granato}(1992)}]{granato92}%
  \BibitemOpen
  \bibfield  {author} {\bibinfo {author} {\bibfnamefont {A.~V.}\ \bibnamefont {Granato}},\ }\bibfield  {title} {\bibinfo {title} {Interstitialcy model for condensed matter states of face-centered-cubic metals},\ }\href {https://doi.org/10.1103/PhysRevLett.68.974} {\bibfield  {journal} {\bibinfo  {journal} {Phys. Rev. Lett.}\ }\textbf {\bibinfo {volume} {68}},\ \bibinfo {pages} {974} (\bibinfo {year} {1992})}\BibitemShut {NoStop}%
\bibitem [{\citenamefont {Boehler}(1986)}]{Boehler86}%
  \BibitemOpen
  \bibfield  {author} {\bibinfo {author} {\bibfnamefont {R.}~\bibnamefont {Boehler}},\ }\bibfield  {title} {\bibinfo {title} {The phase diagram of iron to 430 kbar},\ }\href {https://doi.org/https://doi.org/10.1029/GL013i011p01153} {\bibfield  {journal} {\bibinfo  {journal} {Geophysical Research Letters}\ }\textbf {\bibinfo {volume} {13}},\ \bibinfo {pages} {1153} (\bibinfo {year} {1986})}\BibitemShut {NoStop}%
\bibitem [{\citenamefont {Huguet}\ \emph {et~al.}(2018)\citenamefont {Huguet}, \citenamefont {{Van Orman}}, \citenamefont {Hauck},\ and\ \citenamefont {Willard}}]{HUGUET20189}%
  \BibitemOpen
  \bibfield  {author} {\bibinfo {author} {\bibfnamefont {L.}~\bibnamefont {Huguet}}, \bibinfo {author} {\bibfnamefont {J.~A.}\ \bibnamefont {{Van Orman}}}, \bibinfo {author} {\bibfnamefont {S.~A.}\ \bibnamefont {Hauck}},\ and\ \bibinfo {author} {\bibfnamefont {M.~A.}\ \bibnamefont {Willard}},\ }\bibfield  {title} {\bibinfo {title} {Earth's inner core nucleation paradox},\ }\href {https://doi.org/https://doi.org/10.1016/j.epsl.2018.01.018} {\bibfield  {journal} {\bibinfo  {journal} {Earth and Planetary Science Letters}\ }\textbf {\bibinfo {volume} {487}},\ \bibinfo {pages} {9} (\bibinfo {year} {2018})}\BibitemShut {NoStop}%
\bibitem [{\citenamefont {Hirose}\ \emph {et~al.}(2013)\citenamefont {Hirose}, \citenamefont {Labrosse},\ and\ \citenamefont {Hernlund}}]{Hirose2013}%
  \BibitemOpen
  \bibfield  {author} {\bibinfo {author} {\bibfnamefont {K.}~\bibnamefont {Hirose}}, \bibinfo {author} {\bibfnamefont {S.}~\bibnamefont {Labrosse}},\ and\ \bibinfo {author} {\bibfnamefont {J.}~\bibnamefont {Hernlund}},\ }\bibfield  {title} {\bibinfo {title} {Composition and state of the core},\ }\href {https://doi.org/https://doi.org/10.1146/annurev-earth-050212-124007} {\bibfield  {journal} {\bibinfo  {journal} {Annual Review of Earth and Planetary Sciences}\ }\textbf {\bibinfo {volume} {41}},\ \bibinfo {pages} {657} (\bibinfo {year} {2013})}\BibitemShut {NoStop}%
\bibitem [{\citenamefont {Stacey}(2005)}]{Stacey2005}%
  \BibitemOpen
  \bibfield  {author} {\bibinfo {author} {\bibfnamefont {F.~D.}\ \bibnamefont {Stacey}},\ }\bibfield  {title} {\bibinfo {title} {High pressure equations of state and planetary interiors},\ }\href {https://doi.org/10.1088/0034-4885/68/2/R03} {\bibfield  {journal} {\bibinfo  {journal} {Reports on Progress in Physics}\ }\textbf {\bibinfo {volume} {68}},\ \bibinfo {pages} {341} (\bibinfo {year} {2005})}\BibitemShut {NoStop}%
\bibitem [{\citenamefont {Kuwayama}\ \emph {et~al.}(2020)\citenamefont {Kuwayama}, \citenamefont {Morard}, \citenamefont {Nakajima}, \citenamefont {Hirose}, \citenamefont {Baron}, \citenamefont {Kawaguchi}, \citenamefont {Tsuchiya}, \citenamefont {Ishikawa}, \citenamefont {Hirao},\ and\ \citenamefont {Ohishi}}]{Yasuhiro2020}%
  \BibitemOpen
  \bibfield  {author} {\bibinfo {author} {\bibfnamefont {Y.}~\bibnamefont {Kuwayama}}, \bibinfo {author} {\bibfnamefont {G.}~\bibnamefont {Morard}}, \bibinfo {author} {\bibfnamefont {Y.}~\bibnamefont {Nakajima}}, \bibinfo {author} {\bibfnamefont {K.}~\bibnamefont {Hirose}}, \bibinfo {author} {\bibfnamefont {A.~Q.~R.}\ \bibnamefont {Baron}}, \bibinfo {author} {\bibfnamefont {S.~I.}\ \bibnamefont {Kawaguchi}}, \bibinfo {author} {\bibfnamefont {T.}~\bibnamefont {Tsuchiya}}, \bibinfo {author} {\bibfnamefont {D.}~\bibnamefont {Ishikawa}}, \bibinfo {author} {\bibfnamefont {N.}~\bibnamefont {Hirao}},\ and\ \bibinfo {author} {\bibfnamefont {Y.}~\bibnamefont {Ohishi}},\ }\bibfield  {title} {\bibinfo {title} {Equation of state of liquid iron under extreme conditions},\ }\href {https://doi.org/10.1103/PhysRevLett.124.165701} {\bibfield  {journal} {\bibinfo  {journal} {Phys. Rev. Lett.}\ }\textbf {\bibinfo {volume} {124}},\ \bibinfo {pages} {165701} (\bibinfo {year} {2020})}\BibitemShut {NoStop}%
\bibitem [{\citenamefont {Dorogokupets}\ \emph {et~al.}(2017)\citenamefont {Dorogokupets}, \citenamefont {Dymshits}, \citenamefont {Litasov},\ and\ \citenamefont {Sokolova}}]{Dorogokupets2017}%
  \BibitemOpen
  \bibfield  {author} {\bibinfo {author} {\bibfnamefont {P.~I.}\ \bibnamefont {Dorogokupets}}, \bibinfo {author} {\bibfnamefont {A.~M.}\ \bibnamefont {Dymshits}}, \bibinfo {author} {\bibfnamefont {K.~D.}\ \bibnamefont {Litasov}},\ and\ \bibinfo {author} {\bibfnamefont {T.~S.}\ \bibnamefont {Sokolova}},\ }\bibfield  {title} {\bibinfo {title} {Thermodynamics and equations of state of iron to 350{\thinspace}gpa and 6000{\thinspace}k},\ }\href {https://doi.org/10.1038/srep41863} {\bibfield  {journal} {\bibinfo  {journal} {Scientific Reports}\ }\textbf {\bibinfo {volume} {7}},\ \bibinfo {pages} {41863} (\bibinfo {year} {2017})}\BibitemShut {NoStop}%
\bibitem [{\citenamefont {Zhang}\ and\ \citenamefont {Lin}(2022)}]{zhang22}%
  \BibitemOpen
  \bibfield  {author} {\bibinfo {author} {\bibfnamefont {Y.}~\bibnamefont {Zhang}}\ and\ \bibinfo {author} {\bibfnamefont {J.-F.}\ \bibnamefont {Lin}},\ }\bibfield  {title} {\bibinfo {title} {Molten iron in earth-like exoplanet cores},\ }\href {https://doi.org/10.1126/science.abn2051} {\bibfield  {journal} {\bibinfo  {journal} {Science}\ }\textbf {\bibinfo {volume} {375}},\ \bibinfo {pages} {146} (\bibinfo {year} {2022})},\ \Eprint {https://arxiv.org/abs/https://www.science.org/doi/pdf/10.1126/science.abn2051} {https://www.science.org/doi/pdf/10.1126/science.abn2051} \BibitemShut {NoStop}%
\bibitem [{\citenamefont {Aubert}\ \emph {et~al.}(2013)\citenamefont {Aubert}, \citenamefont {Finlay},\ and\ \citenamefont {Fournier}}]{Aubert2013}%
  \BibitemOpen
  \bibfield  {author} {\bibinfo {author} {\bibfnamefont {J.}~\bibnamefont {Aubert}}, \bibinfo {author} {\bibfnamefont {C.~C.}\ \bibnamefont {Finlay}},\ and\ \bibinfo {author} {\bibfnamefont {A.}~\bibnamefont {Fournier}},\ }\bibfield  {title} {\bibinfo {title} {Bottom-up control of geomagnetic secular variation by the earth's inner core},\ }\href {https://doi.org/10.1038/nature12574} {\bibfield  {journal} {\bibinfo  {journal} {Nature}\ }\textbf {\bibinfo {volume} {502}},\ \bibinfo {pages} {219} (\bibinfo {year} {2013})}\BibitemShut {NoStop}%
\bibitem [{\citenamefont {Landeau}\ \emph {et~al.}(2017)\citenamefont {Landeau}, \citenamefont {Aubert},\ and\ \citenamefont {Olson}}]{LANDEAU2017}%
  \BibitemOpen
  \bibfield  {author} {\bibinfo {author} {\bibfnamefont {M.}~\bibnamefont {Landeau}}, \bibinfo {author} {\bibfnamefont {J.}~\bibnamefont {Aubert}},\ and\ \bibinfo {author} {\bibfnamefont {P.}~\bibnamefont {Olson}},\ }\bibfield  {title} {\bibinfo {title} {The signature of inner-core nucleation on the geodynamo},\ }\href {https://doi.org/https://doi.org/10.1016/j.epsl.2017.02.004} {\bibfield  {journal} {\bibinfo  {journal} {Earth and Planetary Science Letters}\ }\textbf {\bibinfo {volume} {465}},\ \bibinfo {pages} {193} (\bibinfo {year} {2017})}\BibitemShut {NoStop}%
\bibitem [{\citenamefont {Brown}\ and\ \citenamefont {McQueen}(1986)}]{Brown&McQueen}%
  \BibitemOpen
  \bibfield  {author} {\bibinfo {author} {\bibfnamefont {J.~M.}\ \bibnamefont {Brown}}\ and\ \bibinfo {author} {\bibfnamefont {R.~G.}\ \bibnamefont {McQueen}},\ }\bibfield  {title} {\bibinfo {title} {Phase transitions, grüneisen parameter, and elasticity for shocked iron between 77 gpa and 400 gpa},\ }\href {https://doi.org/https://doi.org/10.1029/JB091iB07p07485} {\bibfield  {journal} {\bibinfo  {journal} {Journal of Geophysical Research: Solid Earth}\ }\textbf {\bibinfo {volume} {91}},\ \bibinfo {pages} {7485} (\bibinfo {year} {1986})}\BibitemShut {NoStop}%
\bibitem [{\citenamefont {Yoo}\ \emph {et~al.}(1993)\citenamefont {Yoo}, \citenamefont {Holmes}, \citenamefont {Ross}, \citenamefont {Webb},\ and\ \citenamefont {Pike}}]{Yoo&Holmes}%
  \BibitemOpen
  \bibfield  {author} {\bibinfo {author} {\bibfnamefont {C.~S.}\ \bibnamefont {Yoo}}, \bibinfo {author} {\bibfnamefont {N.~C.}\ \bibnamefont {Holmes}}, \bibinfo {author} {\bibfnamefont {M.}~\bibnamefont {Ross}}, \bibinfo {author} {\bibfnamefont {D.~J.}\ \bibnamefont {Webb}},\ and\ \bibinfo {author} {\bibfnamefont {C.}~\bibnamefont {Pike}},\ }\bibfield  {title} {\bibinfo {title} {Shock temperatures and melting of iron at earth core conditions},\ }\href {https://doi.org/10.1103/PhysRevLett.70.3931} {\bibfield  {journal} {\bibinfo  {journal} {Phys. Rev. Lett.}\ }\textbf {\bibinfo {volume} {70}},\ \bibinfo {pages} {3931} (\bibinfo {year} {1993})}\BibitemShut {NoStop}%
\bibitem [{\citenamefont {Nguyen}\ and\ \citenamefont {Holmes}(2004)}]{Nguyen2004}%
  \BibitemOpen
  \bibfield  {author} {\bibinfo {author} {\bibfnamefont {J.~H.}\ \bibnamefont {Nguyen}}\ and\ \bibinfo {author} {\bibfnamefont {N.~C.}\ \bibnamefont {Holmes}},\ }\bibfield  {title} {\bibinfo {title} {Melting of iron at the physical conditions of the earth's core},\ }\href {https://doi.org/10.1038/nature02248} {\bibfield  {journal} {\bibinfo  {journal} {Nature}\ }\textbf {\bibinfo {volume} {427}},\ \bibinfo {pages} {339} (\bibinfo {year} {2004})}\BibitemShut {NoStop}%
\bibitem [{\citenamefont {Harmand}\ \emph {et~al.}(2015)\citenamefont {Harmand}, \citenamefont {Ravasio}, \citenamefont {Mazevet}, \citenamefont {Bouchet}, \citenamefont {Denoeud}, \citenamefont {Dorchies}, \citenamefont {Feng}, \citenamefont {Fourment}, \citenamefont {Galtier}, \citenamefont {Gaudin}, \citenamefont {Guyot}, \citenamefont {Kodama}, \citenamefont {Koenig}, \citenamefont {Lee}, \citenamefont {Miyanishi}, \citenamefont {Morard}, \citenamefont {Musella}, \citenamefont {Nagler}, \citenamefont {Nakatsutsumi}, \citenamefont {Ozaki}, \citenamefont {Recoules}, \citenamefont {Toleikis}, \citenamefont {Vinci}, \citenamefont {Zastrau}, \citenamefont {Zhu},\ and\ \citenamefont {Benuzzi-Mounaix}}]{Hermand}%
  \BibitemOpen
  \bibfield  {author} {\bibinfo {author} {\bibfnamefont {M.}~\bibnamefont {Harmand}}, \bibinfo {author} {\bibfnamefont {A.}~\bibnamefont {Ravasio}}, \bibinfo {author} {\bibfnamefont {S.}~\bibnamefont {Mazevet}}, \bibinfo {author} {\bibfnamefont {J.}~\bibnamefont {Bouchet}}, \bibinfo {author} {\bibfnamefont {A.}~\bibnamefont {Denoeud}}, \bibinfo {author} {\bibfnamefont {F.}~\bibnamefont {Dorchies}}, \bibinfo {author} {\bibfnamefont {Y.}~\bibnamefont {Feng}}, \bibinfo {author} {\bibfnamefont {C.}~\bibnamefont {Fourment}}, \bibinfo {author} {\bibfnamefont {E.}~\bibnamefont {Galtier}}, \bibinfo {author} {\bibfnamefont {J.}~\bibnamefont {Gaudin}}, \bibinfo {author} {\bibfnamefont {F.}~\bibnamefont {Guyot}}, \bibinfo {author} {\bibfnamefont {R.}~\bibnamefont {Kodama}}, \bibinfo {author} {\bibfnamefont {M.}~\bibnamefont {Koenig}}, \bibinfo {author} {\bibfnamefont {H.~J.}\ \bibnamefont {Lee}}, \bibinfo {author} {\bibfnamefont {K.}~\bibnamefont {Miyanishi}}, \bibinfo {author} {\bibfnamefont {G.}~\bibnamefont
  {Morard}}, \bibinfo {author} {\bibfnamefont {R.}~\bibnamefont {Musella}}, \bibinfo {author} {\bibfnamefont {B.}~\bibnamefont {Nagler}}, \bibinfo {author} {\bibfnamefont {M.}~\bibnamefont {Nakatsutsumi}}, \bibinfo {author} {\bibfnamefont {N.}~\bibnamefont {Ozaki}}, \bibinfo {author} {\bibfnamefont {V.}~\bibnamefont {Recoules}}, \bibinfo {author} {\bibfnamefont {S.}~\bibnamefont {Toleikis}}, \bibinfo {author} {\bibfnamefont {T.}~\bibnamefont {Vinci}}, \bibinfo {author} {\bibfnamefont {U.}~\bibnamefont {Zastrau}}, \bibinfo {author} {\bibfnamefont {D.}~\bibnamefont {Zhu}},\ and\ \bibinfo {author} {\bibfnamefont {A.}~\bibnamefont {Benuzzi-Mounaix}},\ }\bibfield  {title} {\bibinfo {title} {X-ray absorption spectroscopy of iron at multimegabar pressures in laser shock experiments},\ }\href {https://doi.org/10.1103/PhysRevB.92.024108} {\bibfield  {journal} {\bibinfo  {journal} {Phys. Rev. B}\ }\textbf {\bibinfo {volume} {92}},\ \bibinfo {pages} {024108} (\bibinfo {year} {2015})}\BibitemShut {NoStop}%
\bibitem [{\citenamefont {Balugani}\ \emph {et~al.}(2024)\citenamefont {Balugani}, \citenamefont {Hernandez}, \citenamefont {S\'evelin-Radiguet}, \citenamefont {Mathon}, \citenamefont {Recoules}, \citenamefont {Kas}, \citenamefont {Eakins}, \citenamefont {Doyle}, \citenamefont {Ravasio},\ and\ \citenamefont {Torchio}}]{Balugani2024}%
  \BibitemOpen
  \bibfield  {author} {\bibinfo {author} {\bibfnamefont {S.}~\bibnamefont {Balugani}}, \bibinfo {author} {\bibfnamefont {J.~A.}\ \bibnamefont {Hernandez}}, \bibinfo {author} {\bibfnamefont {N.}~\bibnamefont {S\'evelin-Radiguet}}, \bibinfo {author} {\bibfnamefont {O.}~\bibnamefont {Mathon}}, \bibinfo {author} {\bibfnamefont {V.}~\bibnamefont {Recoules}}, \bibinfo {author} {\bibfnamefont {J.~J.}\ \bibnamefont {Kas}}, \bibinfo {author} {\bibfnamefont {D.~E.}\ \bibnamefont {Eakins}}, \bibinfo {author} {\bibfnamefont {H.}~\bibnamefont {Doyle}}, \bibinfo {author} {\bibfnamefont {A.}~\bibnamefont {Ravasio}},\ and\ \bibinfo {author} {\bibfnamefont {R.}~\bibnamefont {Torchio}},\ }\bibfield  {title} {\bibinfo {title} {New constraints on the melting temperature and phase stability of shocked iron up to 270 gpa probed by ultrafast x-ray absorption spectroscopy},\ }\href {https://doi.org/10.1103/PhysRevLett.133.254101} {\bibfield  {journal} {\bibinfo  {journal} {Phys. Rev. Lett.}\ }\textbf {\bibinfo {volume} {133}},\
  \bibinfo {pages} {254101} (\bibinfo {year} {2024})}\BibitemShut {NoStop}%
\bibitem [{\citenamefont {Anzellini}\ \emph {et~al.}(2013)\citenamefont {Anzellini}, \citenamefont {Dewaele}, \citenamefont {Mezouar}, \citenamefont {Loubeyre},\ and\ \citenamefont {Morard}}]{Anzellini2013}%
  \BibitemOpen
  \bibfield  {author} {\bibinfo {author} {\bibfnamefont {S.}~\bibnamefont {Anzellini}}, \bibinfo {author} {\bibfnamefont {A.}~\bibnamefont {Dewaele}}, \bibinfo {author} {\bibfnamefont {M.}~\bibnamefont {Mezouar}}, \bibinfo {author} {\bibfnamefont {P.}~\bibnamefont {Loubeyre}},\ and\ \bibinfo {author} {\bibfnamefont {G.}~\bibnamefont {Morard}},\ }\bibfield  {title} {\bibinfo {title} {Melting of iron at earth’s inner core boundary based on fast x-ray diffraction},\ }\href {https://doi.org/10.1126/science.1233514} {\bibfield  {journal} {\bibinfo  {journal} {Science}\ }\textbf {\bibinfo {volume} {340}},\ \bibinfo {pages} {464} (\bibinfo {year} {2013})}\BibitemShut {NoStop}%
\bibitem [{\citenamefont {Sinmyo}\ \emph {et~al.}(2019)\citenamefont {Sinmyo}, \citenamefont {Hirose},\ and\ \citenamefont {Ohishi}}]{SINMYO2019}%
  \BibitemOpen
  \bibfield  {author} {\bibinfo {author} {\bibfnamefont {R.}~\bibnamefont {Sinmyo}}, \bibinfo {author} {\bibfnamefont {K.}~\bibnamefont {Hirose}},\ and\ \bibinfo {author} {\bibfnamefont {Y.}~\bibnamefont {Ohishi}},\ }\bibfield  {title} {\bibinfo {title} {Melting curve of iron to 290 gpa determined in a resistance-heated diamond-anvil cell},\ }\href {https://doi.org/https://doi.org/10.1016/j.epsl.2019.01.006} {\bibfield  {journal} {\bibinfo  {journal} {Earth and Planetary Science Letters}\ }\textbf {\bibinfo {volume} {510}},\ \bibinfo {pages} {45} (\bibinfo {year} {2019})}\BibitemShut {NoStop}%
\bibitem [{\citenamefont {Boehler}(1993)}]{Boehler1993}%
  \BibitemOpen
  \bibfield  {author} {\bibinfo {author} {\bibfnamefont {R.}~\bibnamefont {Boehler}},\ }\bibfield  {title} {\bibinfo {title} {Temperatures in the earth's core from melting-point measurements of iron at high static pressures},\ }\href {https://doi.org/10.1038/363534a0} {\bibfield  {journal} {\bibinfo  {journal} {Nature}\ }\textbf {\bibinfo {volume} {363}},\ \bibinfo {pages} {534} (\bibinfo {year} {1993})}\BibitemShut {NoStop}%
\bibitem [{\citenamefont {Aquilanti}\ \emph {et~al.}(2015)\citenamefont {Aquilanti}, \citenamefont {Trapananti}, \citenamefont {Karandikar}, \citenamefont {Kantor}, \citenamefont {Marini}, \citenamefont {Mathon}, \citenamefont {Pascarelli},\ and\ \citenamefont {Boehler}}]{Aquilanti2015}%
  \BibitemOpen
  \bibfield  {author} {\bibinfo {author} {\bibfnamefont {G.}~\bibnamefont {Aquilanti}}, \bibinfo {author} {\bibfnamefont {A.}~\bibnamefont {Trapananti}}, \bibinfo {author} {\bibfnamefont {A.}~\bibnamefont {Karandikar}}, \bibinfo {author} {\bibfnamefont {I.}~\bibnamefont {Kantor}}, \bibinfo {author} {\bibfnamefont {C.}~\bibnamefont {Marini}}, \bibinfo {author} {\bibfnamefont {O.}~\bibnamefont {Mathon}}, \bibinfo {author} {\bibfnamefont {S.}~\bibnamefont {Pascarelli}},\ and\ \bibinfo {author} {\bibfnamefont {R.}~\bibnamefont {Boehler}},\ }\bibfield  {title} {\bibinfo {title} {Melting of iron determined by x-ray absorption spectroscopy to 100 gpa},\ }\href {https://doi.org/10.1073/pnas.1502363112} {\bibfield  {journal} {\bibinfo  {journal} {Proceedings of the National Academy of Sciences}\ }\textbf {\bibinfo {volume} {112}},\ \bibinfo {pages} {12042} (\bibinfo {year} {2015})},\ \Eprint {https://arxiv.org/abs/https://www.pnas.org/doi/pdf/10.1073/pnas.1502363112}
  {https://www.pnas.org/doi/pdf/10.1073/pnas.1502363112} \BibitemShut {NoStop}%
\bibitem [{\citenamefont {Tateno}\ \emph {et~al.}(2010)\citenamefont {Tateno}, \citenamefont {Hirose}, \citenamefont {Ohishi},\ and\ \citenamefont {Tatsumi}}]{Tateno2010}%
  \BibitemOpen
  \bibfield  {author} {\bibinfo {author} {\bibfnamefont {S.}~\bibnamefont {Tateno}}, \bibinfo {author} {\bibfnamefont {K.}~\bibnamefont {Hirose}}, \bibinfo {author} {\bibfnamefont {Y.}~\bibnamefont {Ohishi}},\ and\ \bibinfo {author} {\bibfnamefont {Y.}~\bibnamefont {Tatsumi}},\ }\bibfield  {title} {\bibinfo {title} {The structure of iron in earth’s inner core},\ }\href {https://doi.org/10.1126/science.1194662} {\bibfield  {journal} {\bibinfo  {journal} {Science}\ }\textbf {\bibinfo {volume} {330}},\ \bibinfo {pages} {359} (\bibinfo {year} {2010})},\ \Eprint {https://arxiv.org/abs/https://www.science.org/doi/pdf/10.1126/science.1194662} {https://www.science.org/doi/pdf/10.1126/science.1194662} \BibitemShut {NoStop}%
\bibitem [{\citenamefont {Anzellini}\ and\ \citenamefont {Boccato}(2020)}]{Anzellini2020}%
  \BibitemOpen
  \bibfield  {author} {\bibinfo {author} {\bibfnamefont {S.}~\bibnamefont {Anzellini}}\ and\ \bibinfo {author} {\bibfnamefont {S.}~\bibnamefont {Boccato}},\ }\bibfield  {title} {\bibinfo {title} {A practical review of the laser-heated diamond anvil cell for university laboratories and synchrotron applications},\ }\bibfield  {journal} {\bibinfo  {journal} {Crystals}\ }\textbf {\bibinfo {volume} {10}},\ \href {https://doi.org/10.3390/cryst10060459} {10.3390/cryst10060459} (\bibinfo {year} {2020})\BibitemShut {NoStop}%
\bibitem [{\citenamefont {Zel’dovich}\ and\ \citenamefont {Raizer}(2012)}]{Zeldovich}%
  \BibitemOpen
  \bibfield  {author} {\bibinfo {author} {\bibfnamefont {Y.}~\bibnamefont {Zel’dovich}}\ and\ \bibinfo {author} {\bibfnamefont {Y.}~\bibnamefont {Raizer}},\ }\href {https://books.google.co.il/books?id=PULCAgAAQBAJ} {\emph {\bibinfo {title} {Physics of Shock Waves and High-Temperature Hydrodynamic Phenomena}}},\ Dover Books on Physics\ (\bibinfo  {publisher} {Dover Publications},\ \bibinfo {year} {2012})\BibitemShut {NoStop}%
\bibitem [{\citenamefont {Duvall}\ and\ \citenamefont {Graham}(1977)}]{Duvall77}%
  \BibitemOpen
  \bibfield  {author} {\bibinfo {author} {\bibfnamefont {G.~E.}\ \bibnamefont {Duvall}}\ and\ \bibinfo {author} {\bibfnamefont {R.~A.}\ \bibnamefont {Graham}},\ }\bibfield  {title} {\bibinfo {title} {Phase transitions under shock-wave loading},\ }\href {https://doi.org/10.1103/RevModPhys.49.523} {\bibfield  {journal} {\bibinfo  {journal} {Rev. Mod. Phys.}\ }\textbf {\bibinfo {volume} {49}},\ \bibinfo {pages} {523} (\bibinfo {year} {1977})}\BibitemShut {NoStop}%
\bibitem [{\citenamefont {Sola}\ and\ \citenamefont {Alf\`e}(2009)}]{alfePRL}%
  \BibitemOpen
  \bibfield  {author} {\bibinfo {author} {\bibfnamefont {E.}~\bibnamefont {Sola}}\ and\ \bibinfo {author} {\bibfnamefont {D.}~\bibnamefont {Alf\`e}},\ }\bibfield  {title} {\bibinfo {title} {Melting of iron under earth's core conditions from diffusion monte carlo free energy calculations},\ }\href {https://doi.org/10.1103/PhysRevLett.103.078501} {\bibfield  {journal} {\bibinfo  {journal} {Phys. Rev. Lett.}\ }\textbf {\bibinfo {volume} {103}},\ \bibinfo {pages} {078501} (\bibinfo {year} {2009})}\BibitemShut {NoStop}%
\bibitem [{\citenamefont {Alf\`e}(2009)}]{alfePRB}%
  \BibitemOpen
  \bibfield  {author} {\bibinfo {author} {\bibfnamefont {D.}~\bibnamefont {Alf\`e}},\ }\bibfield  {title} {\bibinfo {title} {Temperature of the inner-core boundary of the earth: Melting of iron at high pressure from first-principles coexistence simulations},\ }\href {https://doi.org/10.1103/PhysRevB.79.060101} {\bibfield  {journal} {\bibinfo  {journal} {Phys. Rev. B}\ }\textbf {\bibinfo {volume} {79}},\ \bibinfo {pages} {060101} (\bibinfo {year} {2009})}\BibitemShut {NoStop}%
\bibitem [{\citenamefont {Wu}\ \emph {et~al.}(2023)\citenamefont {Wu}, \citenamefont {Benedict}, \citenamefont {Myint}, \citenamefont {Hamel}, \citenamefont {Prisbrey},\ and\ \citenamefont {Leek}}]{Livermore2023}%
  \BibitemOpen
  \bibfield  {author} {\bibinfo {author} {\bibfnamefont {C.~J.}\ \bibnamefont {Wu}}, \bibinfo {author} {\bibfnamefont {L.~X.}\ \bibnamefont {Benedict}}, \bibinfo {author} {\bibfnamefont {P.~C.}\ \bibnamefont {Myint}}, \bibinfo {author} {\bibfnamefont {S.}~\bibnamefont {Hamel}}, \bibinfo {author} {\bibfnamefont {C.~J.}\ \bibnamefont {Prisbrey}},\ and\ \bibinfo {author} {\bibfnamefont {J.~R.}\ \bibnamefont {Leek}},\ }\bibfield  {title} {\bibinfo {title} {Wide-ranged multiphase equation of state for iron and model variations addressing uncertainties in high-pressure melting},\ }\href {https://doi.org/10.1103/PhysRevB.108.014102} {\bibfield  {journal} {\bibinfo  {journal} {Phys. Rev. B}\ }\textbf {\bibinfo {volume} {108}},\ \bibinfo {pages} {014102} (\bibinfo {year} {2023})}\BibitemShut {NoStop}%
\bibitem [{\citenamefont {Saxena}\ and\ \citenamefont {Eriksson}(2015)}]{saxena_2015}%
  \BibitemOpen
  \bibfield  {author} {\bibinfo {author} {\bibfnamefont {S.~K.}\ \bibnamefont {Saxena}}\ and\ \bibinfo {author} {\bibfnamefont {G.}~\bibnamefont {Eriksson}},\ }\bibfield  {title} {\bibinfo {title} {Thermodynamics of iron at extreme pressures and temperatures},\ }\href {https://doi.org/10.1016/j.jpcs.2015.03.006} {\bibfield  {journal} {\bibinfo  {journal} {Journal of Physics and Chemistry of Solids}\ }\textbf {\bibinfo {volume} {84}},\ \bibinfo {pages} {70} (\bibinfo {year} {2015})}\BibitemShut {NoStop}%
\bibitem [{\citenamefont {Belonoshko}\ and\ \citenamefont {Smirnov}(2023)}]{Smirnov2023}%
  \BibitemOpen
  \bibfield  {author} {\bibinfo {author} {\bibfnamefont {A.~B.}\ \bibnamefont {Belonoshko}}\ and\ \bibinfo {author} {\bibfnamefont {G.~S.}\ \bibnamefont {Smirnov}},\ }\bibfield  {title} {\bibinfo {title} {A comparison of experimental and ab initio structural data on fe under extreme conditions},\ }\bibfield  {journal} {\bibinfo  {journal} {Metals}\ }\textbf {\bibinfo {volume} {13}},\ \href {https://doi.org/10.3390/met13061096} {10.3390/met13061096} (\bibinfo {year} {2023})\BibitemShut {NoStop}%
\bibitem [{\citenamefont {Smirnov}\ \emph {et~al.}(2025)\citenamefont {Smirnov}, \citenamefont {Peil}, \citenamefont {Ruban}, \citenamefont {Simak},\ and\ \citenamefont {Belonoshko}}]{Smirnov2025}%
  \BibitemOpen
  \bibfield  {author} {\bibinfo {author} {\bibfnamefont {G.~S.}\ \bibnamefont {Smirnov}}, \bibinfo {author} {\bibfnamefont {O.~E.}\ \bibnamefont {Peil}}, \bibinfo {author} {\bibfnamefont {A.~V.}\ \bibnamefont {Ruban}}, \bibinfo {author} {\bibfnamefont {S.~I.}\ \bibnamefont {Simak}},\ and\ \bibinfo {author} {\bibfnamefont {A.~B.}\ \bibnamefont {Belonoshko}},\ }\bibfield  {title} {\bibinfo {title} {Impact of magnetism on fe phase diagram under extreme conditions},\ }\href {https://doi.org/10.1103/PhysRevMaterials.9.L040601} {\bibfield  {journal} {\bibinfo  {journal} {Phys. Rev. Mater.}\ }\textbf {\bibinfo {volume} {9}},\ \bibinfo {pages} {L040601} (\bibinfo {year} {2025})}\BibitemShut {NoStop}%
\bibitem [{\citenamefont {Konopkova}\ \emph {et~al.}(2025)\citenamefont {Konopkova}, \citenamefont {Edmund}, \citenamefont {Ball}, \citenamefont {Dewaele}, \citenamefont {Ginestet}, \citenamefont {Husband}, \citenamefont {Jaisle}, \citenamefont {Strohm}, \citenamefont {Anae}, \citenamefont {Antonangeli}, \citenamefont {Appel}, \citenamefont {Baron}, \citenamefont {Boccato}, \citenamefont {Buakor}, \citenamefont {Chantel}, \citenamefont {Cynn}, \citenamefont {Dwivedi}, \citenamefont {Ehm}, \citenamefont {Glazyrin}, \citenamefont {Graafsma}, \citenamefont {Koemets}, \citenamefont {Laurus}, \citenamefont {Marquardt}, \citenamefont {Massani}, \citenamefont {McHardy}, \citenamefont {McMahon}, \citenamefont {Prakapenka}, \citenamefont {Sztuk-Dambietz}, \citenamefont {Tang}, \citenamefont {Xie}, \citenamefont {Younes}, \citenamefont {Zastrau}, \citenamefont {Goncharov}, \citenamefont {Prescher}, \citenamefont {McWilliams}, \citenamefont {Morard},\ and\ \citenamefont {Merkel}}]{konopkova2025}%
  \BibitemOpen
  \bibfield  {author} {\bibinfo {author} {\bibfnamefont {Z.}~\bibnamefont {Konopkova}}, \bibinfo {author} {\bibfnamefont {E.}~\bibnamefont {Edmund}}, \bibinfo {author} {\bibfnamefont {O.~B.}\ \bibnamefont {Ball}}, \bibinfo {author} {\bibfnamefont {A.}~\bibnamefont {Dewaele}}, \bibinfo {author} {\bibfnamefont {H.}~\bibnamefont {Ginestet}}, \bibinfo {author} {\bibfnamefont {R.~J.}\ \bibnamefont {Husband}}, \bibinfo {author} {\bibfnamefont {N.}~\bibnamefont {Jaisle}}, \bibinfo {author} {\bibfnamefont {C.}~\bibnamefont {Strohm}}, \bibinfo {author} {\bibfnamefont {M.~S.}\ \bibnamefont {Anae}}, \bibinfo {author} {\bibfnamefont {D.}~\bibnamefont {Antonangeli}}, \bibinfo {author} {\bibfnamefont {K.}~\bibnamefont {Appel}}, \bibinfo {author} {\bibfnamefont {M.}~\bibnamefont {Baron}}, \bibinfo {author} {\bibfnamefont {S.}~\bibnamefont {Boccato}}, \bibinfo {author} {\bibfnamefont {K.}~\bibnamefont {Buakor}}, \bibinfo {author} {\bibfnamefont {J.}~\bibnamefont {Chantel}}, \bibinfo {author} {\bibfnamefont {H.}~\bibnamefont
  {Cynn}}, \bibinfo {author} {\bibfnamefont {A.~P.}\ \bibnamefont {Dwivedi}}, \bibinfo {author} {\bibfnamefont {L.}~\bibnamefont {Ehm}}, \bibinfo {author} {\bibfnamefont {K.}~\bibnamefont {Glazyrin}}, \bibinfo {author} {\bibfnamefont {H.}~\bibnamefont {Graafsma}}, \bibinfo {author} {\bibfnamefont {E.}~\bibnamefont {Koemets}}, \bibinfo {author} {\bibfnamefont {T.}~\bibnamefont {Laurus}}, \bibinfo {author} {\bibfnamefont {H.}~\bibnamefont {Marquardt}}, \bibinfo {author} {\bibfnamefont {B.}~\bibnamefont {Massani}}, \bibinfo {author} {\bibfnamefont {J.~D.}\ \bibnamefont {McHardy}}, \bibinfo {author} {\bibfnamefont {M.~I.}\ \bibnamefont {McMahon}}, \bibinfo {author} {\bibfnamefont {V.}~\bibnamefont {Prakapenka}}, \bibinfo {author} {\bibfnamefont {J.}~\bibnamefont {Sztuk-Dambietz}}, \bibinfo {author} {\bibfnamefont {M.}~\bibnamefont {Tang}}, \bibinfo {author} {\bibfnamefont {T.}~\bibnamefont {Xie}}, \bibinfo {author} {\bibfnamefont {Z.}~\bibnamefont {Younes}}, \bibinfo {author} {\bibfnamefont {U.}~\bibnamefont
  {Zastrau}}, \bibinfo {author} {\bibfnamefont {A.~F.}\ \bibnamefont {Goncharov}}, \bibinfo {author} {\bibfnamefont {C.}~\bibnamefont {Prescher}}, \bibinfo {author} {\bibfnamefont {R.~S.}\ \bibnamefont {McWilliams}}, \bibinfo {author} {\bibfnamefont {G.}~\bibnamefont {Morard}},\ and\ \bibinfo {author} {\bibfnamefont {S.}~\bibnamefont {Merkel}},\ }\href {https://arxiv.org/abs/2505.15397} {\bibinfo {title} {Observation of body-centered cubic iron above 200 gigapascals}} (\bibinfo {year} {2025}),\ \Eprint {https://arxiv.org/abs/2505.15397} {arXiv:2505.15397 [cond-mat.mtrl-sci]} \BibitemShut {NoStop}%
\bibitem [{\citenamefont {Li}\ and\ \citenamefont {Scandolo}(2024)}]{Li2024}%
  \BibitemOpen
  \bibfield  {author} {\bibinfo {author} {\bibfnamefont {Z.}~\bibnamefont {Li}}\ and\ \bibinfo {author} {\bibfnamefont {S.}~\bibnamefont {Scandolo}},\ }\bibfield  {title} {\bibinfo {title} {Competing phases of iron at earth's core conditions from deep-learning-aided ab-initio simulations},\ }\href {https://doi.org/https://doi.org/10.1029/2024GL110357} {\bibfield  {journal} {\bibinfo  {journal} {Geophysical Research Letters}\ }\textbf {\bibinfo {volume} {51}},\ \bibinfo {pages} {e2024GL110357} (\bibinfo {year} {2024})},\ \bibinfo {note} {e2024GL110357 2024GL110357},\ \Eprint {https://arxiv.org/abs/https://agupubs.onlinelibrary.wiley.com/doi/pdf/10.1029/2024GL110357} {https://agupubs.onlinelibrary.wiley.com/doi/pdf/10.1029/2024GL110357} \BibitemShut {NoStop}%
\bibitem [{\citenamefont {Wallace}(1997)}]{wallace97}%
  \BibitemOpen
  \bibfield  {author} {\bibinfo {author} {\bibfnamefont {D.~C.}\ \bibnamefont {Wallace}},\ }\bibfield  {title} {\bibinfo {title} {Statistical mechanics of monatomic liquids},\ }\href {https://doi.org/10.1103/PhysRevE.56.4179} {\bibfield  {journal} {\bibinfo  {journal} {Phys. Rev. E}\ }\textbf {\bibinfo {volume} {56}},\ \bibinfo {pages} {4179} (\bibinfo {year} {1997})}\BibitemShut {NoStop}%
\bibitem [{\citenamefont {Lennard-Jones}\ and\ \citenamefont {Devonshire}(1939)}]{Lennard-Jones}%
  \BibitemOpen
  \bibfield  {author} {\bibinfo {author} {\bibfnamefont {J.~E.}\ \bibnamefont {Lennard-Jones}}\ and\ \bibinfo {author} {\bibfnamefont {A.~F.}\ \bibnamefont {Devonshire}},\ }\bibfield  {title} {\bibinfo {title} {Critical and co-operative phenomena. iii. a theory of melting and the structure of liquids},\ }\href {https://doi.org/10.1098/rspa.1939.0002} {\bibfield  {journal} {\bibinfo  {journal} {Proceedings of the Royal Society of London. Series A. Mathematical and Physical Sciences}\ }\textbf {\bibinfo {volume} {169}},\ \bibinfo {pages} {317} (\bibinfo {year} {1939})}\BibitemShut {NoStop}%
\bibitem [{\citenamefont {Cotterill}(1980)}]{Cotterill}%
  \BibitemOpen
  \bibfield  {author} {\bibinfo {author} {\bibfnamefont {R.~M.~J.}\ \bibnamefont {Cotterill}},\ }\href@noop {} {\emph {\bibinfo {title} {Ordering in Strongly Fluctuating Condensed Matter Systems}}}\ (\bibinfo  {publisher} {Plenum},\ \bibinfo {address} {New York},\ \bibinfo {year} {1980})\BibitemShut {NoStop}%
\bibitem [{\citenamefont {Kleinert}(1989)}]{Kleinert}%
  \BibitemOpen
  \bibfield  {author} {\bibinfo {author} {\bibfnamefont {H.}~\bibnamefont {Kleinert}},\ }\href {https://doi.org/10.1142/0356} {\emph {\bibinfo {title} {Gauge Fields in Condensed Matter}}}\ (\bibinfo  {publisher} {WORLD SCIENTIFIC},\ \bibinfo {year} {1989})\BibitemShut {NoStop}%
\bibitem [{\citenamefont {de~With}(2023)}]{deWith23}%
  \BibitemOpen
  \bibfield  {author} {\bibinfo {author} {\bibfnamefont {G.}~\bibnamefont {de~With}},\ }\bibfield  {title} {\bibinfo {title} {Melting is well-known, but is it also well-understood?},\ }\href {https://doi.org/10.1021/acs.chemrev.3c00489} {\bibfield  {journal} {\bibinfo  {journal} {Chemical Reviews}\ }\textbf {\bibinfo {volume} {123}},\ \bibinfo {pages} {13713} (\bibinfo {year} {2023})},\ \bibinfo {note} {pMID: 37963286},\ \Eprint {https://arxiv.org/abs/https://doi.org/10.1021/acs.chemrev.3c00489} {https://doi.org/10.1021/acs.chemrev.3c00489} \BibitemShut {NoStop}%
\bibitem [{\citenamefont {Poirier}(1986)}]{Poirier1986}%
  \BibitemOpen
  \bibfield  {author} {\bibinfo {author} {\bibfnamefont {J.~P.}\ \bibnamefont {Poirier}},\ }\bibfield  {title} {\bibinfo {title} {Dislocation-mediated melting of iron and the temperature of the earth's core},\ }\href {https://doi.org/10.1111/j.1365-246X.1986.tb04515.x} {\bibfield  {journal} {\bibinfo  {journal} {Geophysical Journal International}\ }\textbf {\bibinfo {volume} {85}},\ \bibinfo {pages} {315} (\bibinfo {year} {1986})},\ \Eprint {https://arxiv.org/abs/https://academic.oup.com/gji/article-pdf/85/2/315/2159261/85-2-315.pdf} {https://academic.oup.com/gji/article-pdf/85/2/315/2159261/85-2-315.pdf} \BibitemShut {NoStop}%
\bibitem [{\citenamefont {Granato}(1994)}]{granato94}%
  \BibitemOpen
  \bibfield  {author} {\bibinfo {author} {\bibfnamefont {A.}~\bibnamefont {Granato}},\ }\bibfield  {title} {\bibinfo {title} {Self-interstitials as basic structural units of liquids and glasses},\ }\href {https://doi.org/https://doi.org/10.1016/0022-3697(94)90112-0} {\bibfield  {journal} {\bibinfo  {journal} {Journal of Physics and Chemistry of Solids}\ }\textbf {\bibinfo {volume} {55}},\ \bibinfo {pages} {931} (\bibinfo {year} {1994})}\BibitemShut {NoStop}%
\bibitem [{\citenamefont {Granato}(2014)}]{granato2014}%
  \BibitemOpen
  \bibfield  {author} {\bibinfo {author} {\bibfnamefont {A.~V.}\ \bibnamefont {Granato}},\ }\bibfield  {title} {\bibinfo {title} {Interstitialcy theory of simple condensed matter},\ }\href {https://doi.org/10.1140/epjb/e2013-41024-1} {\bibfield  {journal} {\bibinfo  {journal} {The European Physical Journal B}\ }\textbf {\bibinfo {volume} {87}},\ \bibinfo {pages} {18} (\bibinfo {year} {2014})}\BibitemShut {NoStop}%
\bibitem [{\citenamefont {Nordlund}\ \emph {et~al.}(2005)\citenamefont {Nordlund}, \citenamefont {Ashkenazy}, \citenamefont {Averback},\ and\ \citenamefont {Granato}}]{Nordlund2005}%
  \BibitemOpen
  \bibfield  {author} {\bibinfo {author} {\bibfnamefont {K.}~\bibnamefont {Nordlund}}, \bibinfo {author} {\bibfnamefont {Y.}~\bibnamefont {Ashkenazy}}, \bibinfo {author} {\bibfnamefont {R.~S.}\ \bibnamefont {Averback}},\ and\ \bibinfo {author} {\bibfnamefont {A.~V.}\ \bibnamefont {Granato}},\ }\bibfield  {title} {\bibinfo {title} {Strings and interstitials in liquids, glasses and crystals},\ }\href {https://doi.org/10.1209/epl/i2005-10132-1} {\bibfield  {journal} {\bibinfo  {journal} {Europhysics Letters}\ }\textbf {\bibinfo {volume} {71}},\ \bibinfo {pages} {625} (\bibinfo {year} {2005})}\BibitemShut {NoStop}%
\bibitem [{\citenamefont {Ashkenazy}\ and\ \citenamefont {Averback}(2007)}]{Ashkenazy2007}%
  \BibitemOpen
  \bibfield  {author} {\bibinfo {author} {\bibfnamefont {Y.}~\bibnamefont {Ashkenazy}}\ and\ \bibinfo {author} {\bibfnamefont {R.~S.}\ \bibnamefont {Averback}},\ }\bibfield  {title} {\bibinfo {title} {Atomic mechanisms controlling crystallization behaviour in metals at deep undercoolings},\ }\href {https://doi.org/10.1209/0295-5075/79/26005} {\bibfield  {journal} {\bibinfo  {journal} {Europhysics Letters}\ }\textbf {\bibinfo {volume} {79}},\ \bibinfo {pages} {26005} (\bibinfo {year} {2007})}\BibitemShut {NoStop}%
\bibitem [{\citenamefont {Chan}\ \emph {et~al.}(2009)\citenamefont {Chan}, \citenamefont {Averback}, \citenamefont {Cahill},\ and\ \citenamefont {Ashkenazy}}]{Chan2009}%
  \BibitemOpen
  \bibfield  {author} {\bibinfo {author} {\bibfnamefont {W.-L.}\ \bibnamefont {Chan}}, \bibinfo {author} {\bibfnamefont {R.~S.}\ \bibnamefont {Averback}}, \bibinfo {author} {\bibfnamefont {D.~G.}\ \bibnamefont {Cahill}},\ and\ \bibinfo {author} {\bibfnamefont {Y.}~\bibnamefont {Ashkenazy}},\ }\bibfield  {title} {\bibinfo {title} {Solidification velocities in deeply undercooled silver},\ }\href {https://doi.org/10.1103/PhysRevLett.102.095701} {\bibfield  {journal} {\bibinfo  {journal} {Phys. Rev. Lett.}\ }\textbf {\bibinfo {volume} {102}},\ \bibinfo {pages} {095701} (\bibinfo {year} {2009})}\BibitemShut {NoStop}%
\bibitem [{\citenamefont {Wallace}(2002)}]{Wallace}%
  \BibitemOpen
  \bibfield  {author} {\bibinfo {author} {\bibfnamefont {D.}~\bibnamefont {Wallace}},\ }\href {https://books.google.co.il/books?id=lX9qDQAAQBAJ} {\emph {\bibinfo {title} {Statistical Physics of Crystals and Liquids: A Guide to Highly Accurate Equations of State}}},\ G - Reference,Information and Interdisciplinary Subjects Series\ (\bibinfo  {publisher} {World Scientific},\ \bibinfo {year} {2002})\BibitemShut {NoStop}%
\bibitem [{\citenamefont {Bushman}(1992)}]{Bushman}%
  \BibitemOpen
  \bibfield  {author} {\bibinfo {author} {\bibfnamefont {A.}~\bibnamefont {Bushman}},\ }\href {https://books.google.co.il/books?id=xfV9KXCqvjIC} {\emph {\bibinfo {title} {Intense Dynamic Loading Of Condensed Matter}}}\ (\bibinfo  {publisher} {Taylor \& Francis},\ \bibinfo {year} {1992})\BibitemShut {NoStop}%
\bibitem [{\citenamefont {Bushman}\ and\ \citenamefont {Fortov}(1983)}]{Bushman1983}%
  \BibitemOpen
  \bibfield  {author} {\bibinfo {author} {\bibfnamefont {A.~V.}\ \bibnamefont {Bushman}}\ and\ \bibinfo {author} {\bibfnamefont {V.~E.}\ \bibnamefont {Fortov}},\ }\bibfield  {title} {\bibinfo {title} {Model equations of state},\ }\href {https://doi.org/10.1070/PU1983v026n06ABEH004419} {\bibfield  {journal} {\bibinfo  {journal} {Soviet Physics Uspekhi}\ }\textbf {\bibinfo {volume} {26}},\ \bibinfo {pages} {465} (\bibinfo {year} {1983})}\BibitemShut {NoStop}%
\bibitem [{\citenamefont {Vinet}\ \emph {et~al.}(1989)\citenamefont {Vinet}, \citenamefont {Rose}, \citenamefont {Ferrante},\ and\ \citenamefont {Smith}}]{Vinet_1989}%
  \BibitemOpen
  \bibfield  {author} {\bibinfo {author} {\bibfnamefont {P.}~\bibnamefont {Vinet}}, \bibinfo {author} {\bibfnamefont {J.~H.}\ \bibnamefont {Rose}}, \bibinfo {author} {\bibfnamefont {J.}~\bibnamefont {Ferrante}},\ and\ \bibinfo {author} {\bibfnamefont {J.~R.}\ \bibnamefont {Smith}},\ }\bibfield  {title} {\bibinfo {title} {Universal features of the equation of state of solids},\ }\href {https://doi.org/10.1088/0953-8984/1/11/002} {\bibfield  {journal} {\bibinfo  {journal} {Journal of Physics: Condensed Matter}\ }\textbf {\bibinfo {volume} {1}},\ \bibinfo {pages} {1941} (\bibinfo {year} {1989})}\BibitemShut {NoStop}%
\bibitem [{\citenamefont {Debye}(1912)}]{Debye1912}%
  \BibitemOpen
  \bibfield  {author} {\bibinfo {author} {\bibfnamefont {P.}~\bibnamefont {Debye}},\ }\bibfield  {title} {\bibinfo {title} {Zur theorie der spezifischen wärmen},\ }\href {https://doi.org/https://doi.org/10.1002/andp.19123441404} {\bibfield  {journal} {\bibinfo  {journal} {Annalen der Physik}\ }\textbf {\bibinfo {volume} {344}},\ \bibinfo {pages} {789} (\bibinfo {year} {1912})}\BibitemShut {NoStop}%
\bibitem [{\citenamefont {Abdullaev}\ \emph {et~al.}(2020)\citenamefont {Abdullaev}, \citenamefont {Khairulin},\ and\ \citenamefont {Stankus}}]{Abdullaev_2020}%
  \BibitemOpen
  \bibfield  {author} {\bibinfo {author} {\bibfnamefont {R.~N.}\ \bibnamefont {Abdullaev}}, \bibinfo {author} {\bibfnamefont {R.~A.}\ \bibnamefont {Khairulin}},\ and\ \bibinfo {author} {\bibfnamefont {S.~V.}\ \bibnamefont {Stankus}},\ }\bibfield  {title} {\bibinfo {title} {Volumetric properties of iron in the solid and liquid states},\ }\href {https://doi.org/10.1088/1742-6596/1675/1/012087} {\bibfield  {journal} {\bibinfo  {journal} {Journal of Physics: Conference Series}\ }\textbf {\bibinfo {volume} {1675}},\ \bibinfo {pages} {012087} (\bibinfo {year} {2020})}\BibitemShut {NoStop}%
\bibitem [{\citenamefont {Assael}\ \emph {et~al.}(2006)\citenamefont {Assael}, \citenamefont {Kakosimos}, \citenamefont {Banish}, \citenamefont {Brillo}, \citenamefont {Egry}, \citenamefont {Brooks}, \citenamefont {Quested}, \citenamefont {Mills}, \citenamefont {Nagashima}, \citenamefont {Sato},\ and\ \citenamefont {Wakeham}}]{Assael2006}%
  \BibitemOpen
  \bibfield  {author} {\bibinfo {author} {\bibfnamefont {M.~J.}\ \bibnamefont {Assael}}, \bibinfo {author} {\bibfnamefont {K.}~\bibnamefont {Kakosimos}}, \bibinfo {author} {\bibfnamefont {R.~M.}\ \bibnamefont {Banish}}, \bibinfo {author} {\bibfnamefont {J.}~\bibnamefont {Brillo}}, \bibinfo {author} {\bibfnamefont {I.}~\bibnamefont {Egry}}, \bibinfo {author} {\bibfnamefont {R.}~\bibnamefont {Brooks}}, \bibinfo {author} {\bibfnamefont {P.~N.}\ \bibnamefont {Quested}}, \bibinfo {author} {\bibfnamefont {K.~C.}\ \bibnamefont {Mills}}, \bibinfo {author} {\bibfnamefont {A.}~\bibnamefont {Nagashima}}, \bibinfo {author} {\bibfnamefont {Y.}~\bibnamefont {Sato}},\ and\ \bibinfo {author} {\bibfnamefont {W.~A.}\ \bibnamefont {Wakeham}},\ }\bibfield  {title} {\bibinfo {title} {{Reference Data for the Density and Viscosity of Liquid Aluminum and Liquid Iron}},\ }\href {https://doi.org/10.1063/1.2149380} {\bibfield  {journal} {\bibinfo  {journal} {Journal of Physical and Chemical Reference Data}\ }\textbf {\bibinfo {volume}
  {35}},\ \bibinfo {pages} {285} (\bibinfo {year} {2006})}\BibitemShut {NoStop}%
\bibitem [{\citenamefont {Chase}(1998)}]{NIST_JANAF}%
  \BibitemOpen
  \bibfield  {author} {\bibinfo {author} {\bibfnamefont {M.}~\bibnamefont {Chase}},\ }\href@noop {} {\emph {\bibinfo {title} {NIST-JANAF Thermochemical Tables, 4th Edition}}}\ (\bibinfo  {publisher} {American Institute of Physics, -1},\ \bibinfo {year} {1998})\BibitemShut {NoStop}%
\bibitem [{\citenamefont {Dewaele}\ \emph {et~al.}(2008)\citenamefont {Dewaele}, \citenamefont {Torrent}, \citenamefont {Loubeyre},\ and\ \citenamefont {Mezouar}}]{Dewaele2008}%
  \BibitemOpen
  \bibfield  {author} {\bibinfo {author} {\bibfnamefont {A.}~\bibnamefont {Dewaele}}, \bibinfo {author} {\bibfnamefont {M.}~\bibnamefont {Torrent}}, \bibinfo {author} {\bibfnamefont {P.}~\bibnamefont {Loubeyre}},\ and\ \bibinfo {author} {\bibfnamefont {M.}~\bibnamefont {Mezouar}},\ }\bibfield  {title} {\bibinfo {title} {Compression curves of transition metals in the mbar range: Experiments and projector augmented-wave calculations},\ }\href {https://doi.org/10.1103/PhysRevB.78.104102} {\bibfield  {journal} {\bibinfo  {journal} {Phys. Rev. B}\ }\textbf {\bibinfo {volume} {78}},\ \bibinfo {pages} {104102} (\bibinfo {year} {2008})}\BibitemShut {NoStop}%
\bibitem [{\citenamefont {Barker}\ and\ \citenamefont {Hollenbach}(1974)}]{Barker1974}%
  \BibitemOpen
  \bibfield  {author} {\bibinfo {author} {\bibfnamefont {L.~M.}\ \bibnamefont {Barker}}\ and\ \bibinfo {author} {\bibfnamefont {R.~E.}\ \bibnamefont {Hollenbach}},\ }\bibfield  {title} {\bibinfo {title} {{Shock wave study of the $\alpha \rightleftharpoons \varepsilon$ phase transition in iron}},\ }\href {https://doi.org/10.1063/1.1663148} {\bibfield  {journal} {\bibinfo  {journal} {Journal of Applied Physics}\ }\textbf {\bibinfo {volume} {45}},\ \bibinfo {pages} {4872} (\bibinfo {year} {1974})}\BibitemShut {NoStop}%
\bibitem [{\citenamefont {Brown}\ \emph {et~al.}(2000)\citenamefont {Brown}, \citenamefont {Fritz},\ and\ \citenamefont {Hixson}}]{Brown2000}%
  \BibitemOpen
  \bibfield  {author} {\bibinfo {author} {\bibfnamefont {J.~M.}\ \bibnamefont {Brown}}, \bibinfo {author} {\bibfnamefont {J.~N.}\ \bibnamefont {Fritz}},\ and\ \bibinfo {author} {\bibfnamefont {R.~S.}\ \bibnamefont {Hixson}},\ }\bibfield  {title} {\bibinfo {title} {{Hugoniot data for iron}},\ }\href {https://doi.org/10.1063/1.1319320} {\bibfield  {journal} {\bibinfo  {journal} {Journal of Applied Physics}\ }\textbf {\bibinfo {volume} {88}},\ \bibinfo {pages} {5496} (\bibinfo {year} {2000})}\BibitemShut {NoStop}%
\bibitem [{\citenamefont {Chen}\ and\ \citenamefont {Ahrens}(1997)}]{Chen1997}%
  \BibitemOpen
  \bibfield  {author} {\bibinfo {author} {\bibfnamefont {G.~Q.}\ \bibnamefont {Chen}}\ and\ \bibinfo {author} {\bibfnamefont {T.~J.}\ \bibnamefont {Ahrens}},\ }\bibfield  {title} {\bibinfo {title} {High pressure and high temperature equation-of-state of gamma and liquid iron},\ }\href {https://doi.org/10.1557/PROC-499-41} {\bibfield  {journal} {\bibinfo  {journal} {MRS Online Proceedings Library}\ }\textbf {\bibinfo {volume} {499}},\ \bibinfo {pages} {41} (\bibinfo {year} {1997})}\BibitemShut {NoStop}%
\bibitem [{\citenamefont {Smith}\ \emph {et~al.}(2018)\citenamefont {Smith}, \citenamefont {Fratanduono}, \citenamefont {Braun}, \citenamefont {Duffy}, \citenamefont {Wicks}, \citenamefont {Celliers}, \citenamefont {Ali}, \citenamefont {Fernandez-Pa{\~{n}}ella}, \citenamefont {Kraus}, \citenamefont {Swift}, \citenamefont {Collins},\ and\ \citenamefont {Eggert}}]{Smith2018}%
  \BibitemOpen
  \bibfield  {author} {\bibinfo {author} {\bibfnamefont {R.~F.}\ \bibnamefont {Smith}}, \bibinfo {author} {\bibfnamefont {D.~E.}\ \bibnamefont {Fratanduono}}, \bibinfo {author} {\bibfnamefont {D.~G.}\ \bibnamefont {Braun}}, \bibinfo {author} {\bibfnamefont {T.~S.}\ \bibnamefont {Duffy}}, \bibinfo {author} {\bibfnamefont {J.~K.}\ \bibnamefont {Wicks}}, \bibinfo {author} {\bibfnamefont {P.~M.}\ \bibnamefont {Celliers}}, \bibinfo {author} {\bibfnamefont {S.~J.}\ \bibnamefont {Ali}}, \bibinfo {author} {\bibfnamefont {A.}~\bibnamefont {Fernandez-Pa{\~{n}}ella}}, \bibinfo {author} {\bibfnamefont {R.~G.}\ \bibnamefont {Kraus}}, \bibinfo {author} {\bibfnamefont {D.~C.}\ \bibnamefont {Swift}}, \bibinfo {author} {\bibfnamefont {G.~W.}\ \bibnamefont {Collins}},\ and\ \bibinfo {author} {\bibfnamefont {J.~H.}\ \bibnamefont {Eggert}},\ }\bibfield  {title} {\bibinfo {title} {Equation of state of iron under core conditions of large rocky exoplanets},\ }\href {https://doi.org/10.1038/s41550-018-0437-9} {\bibfield  {journal}
  {\bibinfo  {journal} {Nature Astronomy}\ }\textbf {\bibinfo {volume} {2}},\ \bibinfo {pages} {452} (\bibinfo {year} {2018})}\BibitemShut {NoStop}%
\bibitem [{\citenamefont {Boettger}\ and\ \citenamefont {Wallace}(1997)}]{Boettgaer1997}%
  \BibitemOpen
  \bibfield  {author} {\bibinfo {author} {\bibfnamefont {J.~C.}\ \bibnamefont {Boettger}}\ and\ \bibinfo {author} {\bibfnamefont {D.~C.}\ \bibnamefont {Wallace}},\ }\bibfield  {title} {\bibinfo {title} {Metastability and dynamics of the shock-induced phase transition in iron},\ }\href {https://doi.org/10.1103/PhysRevB.55.2840} {\bibfield  {journal} {\bibinfo  {journal} {Phys. Rev. B}\ }\textbf {\bibinfo {volume} {55}},\ \bibinfo {pages} {2840} (\bibinfo {year} {1997})}\BibitemShut {NoStop}%
\bibitem [{\citenamefont {Swartzendruber}(1982)}]{Swartzendruber1982}%
  \BibitemOpen
  \bibfield  {author} {\bibinfo {author} {\bibfnamefont {L.~J.}\ \bibnamefont {Swartzendruber}},\ }\bibfield  {title} {\bibinfo {title} {The fe (iron) system},\ }\href {https://doi.org/10.1007/BF02892374} {\bibfield  {journal} {\bibinfo  {journal} {Bulletin of Alloy Phase Diagrams}\ }\textbf {\bibinfo {volume} {3}},\ \bibinfo {pages} {161} (\bibinfo {year} {1982})}\BibitemShut {NoStop}%
\bibitem [{\citenamefont {Tiwari}(1978)}]{richards_rule}%
  \BibitemOpen
  \bibfield  {author} {\bibinfo {author} {\bibfnamefont {G.~P.}\ \bibnamefont {Tiwari}},\ }\bibfield  {title} {\bibinfo {title} {Modification of richard's rule and correlation between entropy of fusion and allotropic behaviour},\ }\href {https://doi.org/10.1179/msc.1978.12.7.317} {\bibfield  {journal} {\bibinfo  {journal} {Metal Science}\ }\textbf {\bibinfo {volume} {12}},\ \bibinfo {pages} {317} (\bibinfo {year} {1978})}\BibitemShut {NoStop}%
\bibitem [{\citenamefont {Ai}\ \emph {et~al.}(2020)\citenamefont {Ai}, \citenamefont {Long}, \citenamefont {Zhang}, \citenamefont {Chen}, \citenamefont {Dong}, \citenamefont {Liu}, \citenamefont {Zhang},\ and\ \citenamefont {Duan}}]{Ai2020}%
  \BibitemOpen
  \bibfield  {author} {\bibinfo {author} {\bibfnamefont {S.}~\bibnamefont {Ai}}, \bibinfo {author} {\bibfnamefont {M.}~\bibnamefont {Long}}, \bibinfo {author} {\bibfnamefont {S.}~\bibnamefont {Zhang}}, \bibinfo {author} {\bibfnamefont {D.}~\bibnamefont {Chen}}, \bibinfo {author} {\bibfnamefont {Z.}~\bibnamefont {Dong}}, \bibinfo {author} {\bibfnamefont {P.}~\bibnamefont {Liu}}, \bibinfo {author} {\bibfnamefont {Y.}~\bibnamefont {Zhang}},\ and\ \bibinfo {author} {\bibfnamefont {H.}~\bibnamefont {Duan}},\ }\href {https://doi.org/10.3390/met10020283} {\bibinfo {title} {Ab initio calculations on elastic properties of if steel matrix phase at high temperature based on lattice expansion theory}} (\bibinfo {year} {2020})\BibitemShut {NoStop}%
\bibitem [{\citenamefont {Cohen}\ \emph {et~al.}(1997)\citenamefont {Cohen}, \citenamefont {Stixrude},\ and\ \citenamefont {Wasserman}}]{coehn1997}%
  \BibitemOpen
  \bibfield  {author} {\bibinfo {author} {\bibfnamefont {R.~E.}\ \bibnamefont {Cohen}}, \bibinfo {author} {\bibfnamefont {L.}~\bibnamefont {Stixrude}},\ and\ \bibinfo {author} {\bibfnamefont {E.}~\bibnamefont {Wasserman}},\ }\bibfield  {title} {\bibinfo {title} {Tight-binding computations of elastic anisotropy of fe, xe, and si under compression},\ }\href {https://doi.org/10.1103/PhysRevB.56.8575} {\bibfield  {journal} {\bibinfo  {journal} {Phys. Rev. B}\ }\textbf {\bibinfo {volume} {56}},\ \bibinfo {pages} {8575} (\bibinfo {year} {1997})}\BibitemShut {NoStop}%
\bibitem [{\citenamefont {S\"oderlind}\ \emph {et~al.}(1996)\citenamefont {S\"oderlind}, \citenamefont {Moriarty},\ and\ \citenamefont {Wills}}]{soderlind1996}%
  \BibitemOpen
  \bibfield  {author} {\bibinfo {author} {\bibfnamefont {P.}~\bibnamefont {S\"oderlind}}, \bibinfo {author} {\bibfnamefont {J.~A.}\ \bibnamefont {Moriarty}},\ and\ \bibinfo {author} {\bibfnamefont {J.~M.}\ \bibnamefont {Wills}},\ }\bibfield  {title} {\bibinfo {title} {First-principles theory of iron up to earth-core pressures: Structural, vibrational, and elastic properties},\ }\href {https://doi.org/10.1103/PhysRevB.53.14063} {\bibfield  {journal} {\bibinfo  {journal} {Phys. Rev. B}\ }\textbf {\bibinfo {volume} {53}},\ \bibinfo {pages} {14063} (\bibinfo {year} {1996})}\BibitemShut {NoStop}%
\bibitem [{\citenamefont {McMillan}(2000)}]{McMillan2000}%
  \BibitemOpen
  \bibfield  {author} {\bibinfo {author} {\bibfnamefont {P.}~\bibnamefont {McMillan}},\ }\bibfield  {title} {\bibinfo {title} {Jumping between liquid states},\ }\href {https://doi.org/10.1038/35003088} {\bibfield  {journal} {\bibinfo  {journal} {Nature}\ }\textbf {\bibinfo {volume} {403}},\ \bibinfo {pages} {151} (\bibinfo {year} {2000})}\BibitemShut {NoStop}%
\bibitem [{\citenamefont {Franzese}\ \emph {et~al.}(2001)\citenamefont {Franzese}, \citenamefont {Malescio}, \citenamefont {Skibinsky}, \citenamefont {Buldyrev},\ and\ \citenamefont {Stanley}}]{Franzese2001}%
  \BibitemOpen
  \bibfield  {author} {\bibinfo {author} {\bibfnamefont {G.}~\bibnamefont {Franzese}}, \bibinfo {author} {\bibfnamefont {G.}~\bibnamefont {Malescio}}, \bibinfo {author} {\bibfnamefont {A.}~\bibnamefont {Skibinsky}}, \bibinfo {author} {\bibfnamefont {S.~V.}\ \bibnamefont {Buldyrev}},\ and\ \bibinfo {author} {\bibfnamefont {H.~E.}\ \bibnamefont {Stanley}},\ }\bibfield  {title} {\bibinfo {title} {Generic mechanism for generating a liquid--liquid phase transition},\ }\href {https://doi.org/10.1038/35055514} {\bibfield  {journal} {\bibinfo  {journal} {Nature}\ }\textbf {\bibinfo {volume} {409}},\ \bibinfo {pages} {692} (\bibinfo {year} {2001})}\BibitemShut {NoStop}%
\bibitem [{\citenamefont {Kraus}\ \emph {et~al.}(2022)\citenamefont {Kraus}, \citenamefont {Hemley}, \citenamefont {Ali}, \citenamefont {Belof}, \citenamefont {Benedict}, \citenamefont {Bernier}, \citenamefont {Braun}, \citenamefont {Cohen}, \citenamefont {Collins}, \citenamefont {Coppari}, \citenamefont {Desjarlais}, \citenamefont {Fratanduono}, \citenamefont {Hamel}, \citenamefont {Krygier}, \citenamefont {Lazicki}, \citenamefont {Mcnaney}, \citenamefont {Millot}, \citenamefont {Myint}, \citenamefont {Newman}, \citenamefont {Rygg}, \citenamefont {Sterbentz}, \citenamefont {Stewart}, \citenamefont {Stixrude}, \citenamefont {Swift}, \citenamefont {Wehrenberg},\ and\ \citenamefont {Eggert}}]{Kraus2022}%
  \BibitemOpen
  \bibfield  {author} {\bibinfo {author} {\bibfnamefont {R.~G.}\ \bibnamefont {Kraus}}, \bibinfo {author} {\bibfnamefont {R.~J.}\ \bibnamefont {Hemley}}, \bibinfo {author} {\bibfnamefont {S.~J.}\ \bibnamefont {Ali}}, \bibinfo {author} {\bibfnamefont {J.~L.}\ \bibnamefont {Belof}}, \bibinfo {author} {\bibfnamefont {L.~X.}\ \bibnamefont {Benedict}}, \bibinfo {author} {\bibfnamefont {J.}~\bibnamefont {Bernier}}, \bibinfo {author} {\bibfnamefont {D.}~\bibnamefont {Braun}}, \bibinfo {author} {\bibfnamefont {R.~E.}\ \bibnamefont {Cohen}}, \bibinfo {author} {\bibfnamefont {G.~W.}\ \bibnamefont {Collins}}, \bibinfo {author} {\bibfnamefont {F.}~\bibnamefont {Coppari}}, \bibinfo {author} {\bibfnamefont {M.~P.}\ \bibnamefont {Desjarlais}}, \bibinfo {author} {\bibfnamefont {D.}~\bibnamefont {Fratanduono}}, \bibinfo {author} {\bibfnamefont {S.}~\bibnamefont {Hamel}}, \bibinfo {author} {\bibfnamefont {A.}~\bibnamefont {Krygier}}, \bibinfo {author} {\bibfnamefont {A.}~\bibnamefont {Lazicki}}, \bibinfo {author}
  {\bibfnamefont {J.}~\bibnamefont {Mcnaney}}, \bibinfo {author} {\bibfnamefont {M.}~\bibnamefont {Millot}}, \bibinfo {author} {\bibfnamefont {P.~C.}\ \bibnamefont {Myint}}, \bibinfo {author} {\bibfnamefont {M.~G.}\ \bibnamefont {Newman}}, \bibinfo {author} {\bibfnamefont {J.~R.}\ \bibnamefont {Rygg}}, \bibinfo {author} {\bibfnamefont {D.~M.}\ \bibnamefont {Sterbentz}}, \bibinfo {author} {\bibfnamefont {S.~T.}\ \bibnamefont {Stewart}}, \bibinfo {author} {\bibfnamefont {L.}~\bibnamefont {Stixrude}}, \bibinfo {author} {\bibfnamefont {D.~C.}\ \bibnamefont {Swift}}, \bibinfo {author} {\bibfnamefont {C.}~\bibnamefont {Wehrenberg}},\ and\ \bibinfo {author} {\bibfnamefont {J.~H.}\ \bibnamefont {Eggert}},\ }\bibfield  {title} {\bibinfo {title} {Measuring the melting curve of iron at super-earth core conditions},\ }\href {https://doi.org/10.1126/science.abm1472} {\bibfield  {journal} {\bibinfo  {journal} {Science}\ }\textbf {\bibinfo {volume} {375}},\ \bibinfo {pages} {202} (\bibinfo {year} {2022})},\ \Eprint
  {https://arxiv.org/abs/https://www.science.org/doi/pdf/10.1126/science.abm1472} {https://www.science.org/doi/pdf/10.1126/science.abm1472} \BibitemShut {NoStop}%
\bibitem [{\citenamefont {Zhang}\ \emph {et~al.}(2023)\citenamefont {Zhang}, \citenamefont {Wang}, \citenamefont {Huang}, \citenamefont {Wang}, \citenamefont {Liang}, \citenamefont {Hao}, \citenamefont {Gao}, \citenamefont {Li}, \citenamefont {Wu}, \citenamefont {Zhang}, \citenamefont {Liu}, \citenamefont {Sun},\ and\ \citenamefont {Lin}}]{Zhang2023}%
  \BibitemOpen
  \bibfield  {author} {\bibinfo {author} {\bibfnamefont {Y.}~\bibnamefont {Zhang}}, \bibinfo {author} {\bibfnamefont {Y.}~\bibnamefont {Wang}}, \bibinfo {author} {\bibfnamefont {Y.}~\bibnamefont {Huang}}, \bibinfo {author} {\bibfnamefont {J.}~\bibnamefont {Wang}}, \bibinfo {author} {\bibfnamefont {Z.}~\bibnamefont {Liang}}, \bibinfo {author} {\bibfnamefont {L.}~\bibnamefont {Hao}}, \bibinfo {author} {\bibfnamefont {Z.}~\bibnamefont {Gao}}, \bibinfo {author} {\bibfnamefont {J.}~\bibnamefont {Li}}, \bibinfo {author} {\bibfnamefont {Q.}~\bibnamefont {Wu}}, \bibinfo {author} {\bibfnamefont {H.}~\bibnamefont {Zhang}}, \bibinfo {author} {\bibfnamefont {Y.}~\bibnamefont {Liu}}, \bibinfo {author} {\bibfnamefont {J.}~\bibnamefont {Sun}},\ and\ \bibinfo {author} {\bibfnamefont {J.-F.}\ \bibnamefont {Lin}},\ }\bibfield  {title} {\bibinfo {title} {Collective motion in hcp-fe at earth’s inner core conditions},\ }\href {https://doi.org/10.1073/pnas.2309952120} {\bibfield  {journal} {\bibinfo  {journal} {Proceedings of
  the National Academy of Sciences}\ }\textbf {\bibinfo {volume} {120}},\ \bibinfo {pages} {e2309952120} (\bibinfo {year} {2023})},\ \Eprint {https://arxiv.org/abs/https://www.pnas.org/doi/pdf/10.1073/pnas.2309952120} {https://www.pnas.org/doi/pdf/10.1073/pnas.2309952120} \BibitemShut {NoStop}%
\bibitem [{\citenamefont {{Sun}}\ \emph {et~al.}(2021)\citenamefont {{Sun}}, \citenamefont {{Zhang}}, \citenamefont {{Mendelev}}, \citenamefont {{Wentzcovitch}},\ and\ \citenamefont {{Ho}}}]{Two_step}%
  \BibitemOpen
  \bibfield  {author} {\bibinfo {author} {\bibfnamefont {Y.}~\bibnamefont {{Sun}}}, \bibinfo {author} {\bibfnamefont {F.}~\bibnamefont {{Zhang}}}, \bibinfo {author} {\bibfnamefont {M.}~\bibnamefont {{Mendelev}}}, \bibinfo {author} {\bibfnamefont {R.}~\bibnamefont {{Wentzcovitch}}},\ and\ \bibinfo {author} {\bibfnamefont {K.~M.}\ \bibnamefont {{Ho}}},\ }\bibfield  {title} {\bibinfo {title} {{Two-step nucleation of the Earth's inner core}},\ }in\ \href@noop {} {\emph {\bibinfo {booktitle} {AGU Fall Meeting Abstracts}}},\ Vol.\ \bibinfo {volume} {2021}\ (\bibinfo {year} {2021})\ pp.\ \bibinfo {pages} {DI35D--0058}\BibitemShut {NoStop}%
\bibitem [{\citenamefont {Shikler}\ and\ \citenamefont {Ashkenazy}(2025)}]{dataset}%
  \BibitemOpen
  \bibfield  {author} {\bibinfo {author} {\bibfnamefont {R.}~\bibnamefont {Shikler}}\ and\ \bibinfo {author} {\bibfnamefont {Y.}~\bibnamefont {Ashkenazy}},\ }\href {https://doi.org/10.7910/DVN/NA3S2H} {\bibinfo {title} {{Data for ``Model for the Thermodynamics of Iron at High Pressures Near Melting''}}} (\bibinfo {year} {2025}),\ \Eprint {https://arxiv.org/abs/https://doi.org/10.7910/DVN/NA3S2H} {https://doi.org/10.7910/DVN/NA3S2H} \BibitemShut {NoStop}%
\end{thebibliography}%

\clearpage
\onecolumngrid  

\appendix

\section{Solid phases parameterization}
\label{Appendi:Solid}

Here we detail the parameters of the free energy functions for the solid  ($\gamma(FCC)$ and $\varepsilon(HCP)$) phases. These parameters were obtained by fitting the Helmholtz free energies described in the main text, equations 5 to 8. 

\begin{table}[H]    
    \centering
    \caption{Solid fitting parameters}
    \begin{ruledtabular}
    \begin{tabular}{cccccccc}
         Phase&  $E_{0}[\text{erg}]$ & $B_{0}[\text{GPa}]$ & $B'$ & $\rho_{0}[\frac{\text{g}}{\text{cc}}]$ & $C_{v}[\frac{\text{erg}}{\text{K}}]$ & $\beta_{e}\cdot\rho_{0}^{\gamma_{e}}$ & $\gamma_{e}$ \\
         $\gamma(FCC)$& $8.37\cdot10^{8}$ & 130.6 & 7.34 & 8.226 & $1.4\cdot10^{6}$ & $1.7\cdot10^{7}$ & 2.86\\
         $\varepsilon(HCP)$& $8.8\cdot10^{8}$ & 182.9 & 5.56 & 8.445 & $1.64\cdot10^{6}$ & $1.1\cdot10^{7}$ & 3.33
    \label{tab:solid}
    \end{tabular}
    \end{ruledtabular}
    
\end{table}

The Debye temperature $\theta(\rho)$ described in [7] was fitted using a fourth-order polynomial, as shown in Table~\ref{tab:debye}.

\begin{table}[H]
    \centering
    \caption{Debye temperature $\theta (\rho) = \Sigma_{0}^{n} c_{i}\rho^{i}$}
    \begin{ruledtabular}
    \begin{tabular}{cccccc}
         Phase&  $c_{0}$ & $c_{1}$ & $c_{2}$ & $c_{3}$ & $c_{4}$ \\
         $\gamma(FCC)$& -41.739 & 141.568 & -23.36 & 1.863 & -0.0365 \\
         $\varepsilon(HCP)$& -39.845 & 147.571 & -25.602 & 1.805 & -0.0307
    \end{tabular}
    \end{ruledtabular}
    
    \label{tab:debye}
\end{table}

\begin{figure}[H]
    \centering
    \begin{subfigure}[t]{0.48\textwidth}
        \includegraphics[width=\textwidth]{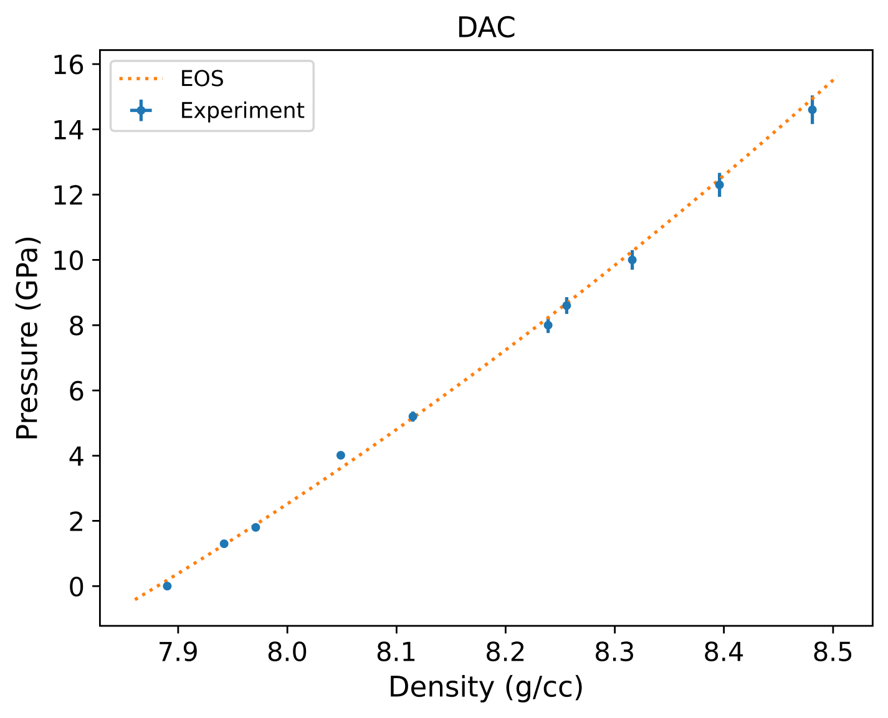}
        \caption*{(a)}
    \end{subfigure}
    \hfill
    \begin{subfigure}[t]{0.48\textwidth}
        \includegraphics[width=\textwidth]{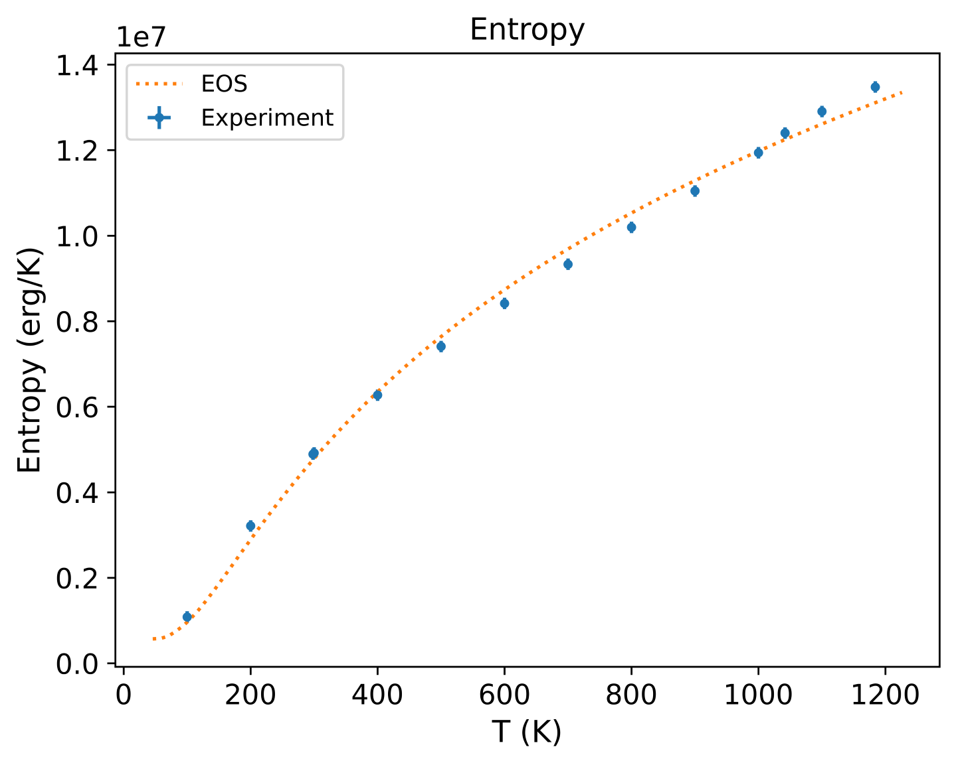}
        \caption*{(b)}
    \end{subfigure}
    
    \vspace{1em}
    
    \begin{subfigure}[t]{0.48\textwidth}
        \includegraphics[width=\textwidth]{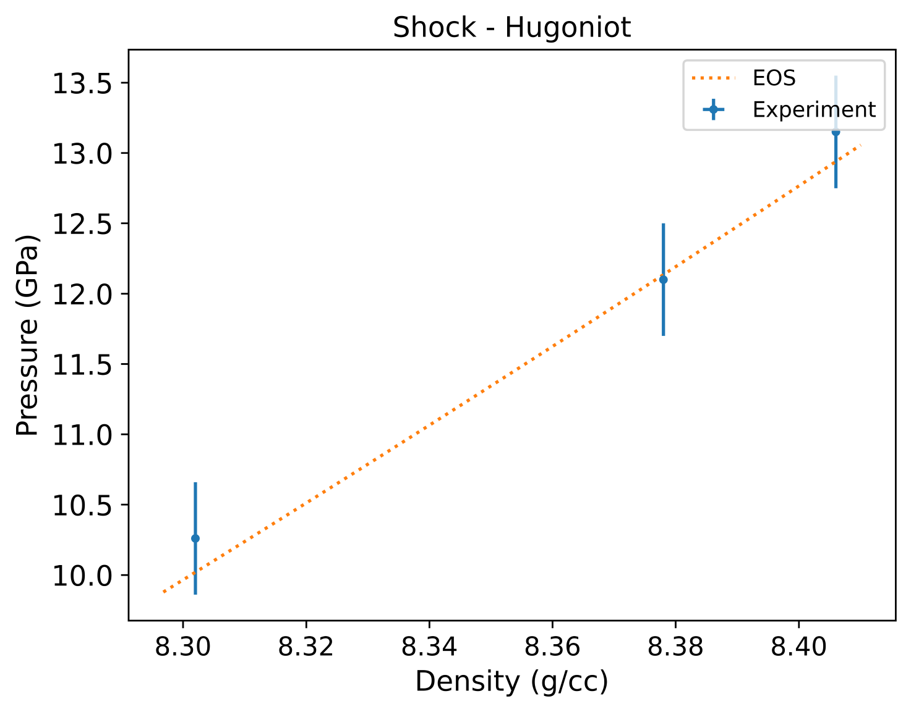}
        \caption*{(c)}
    \end{subfigure}
    \hfill
    \begin{subfigure}[t]{0.48\textwidth}
        \includegraphics[width=\textwidth]{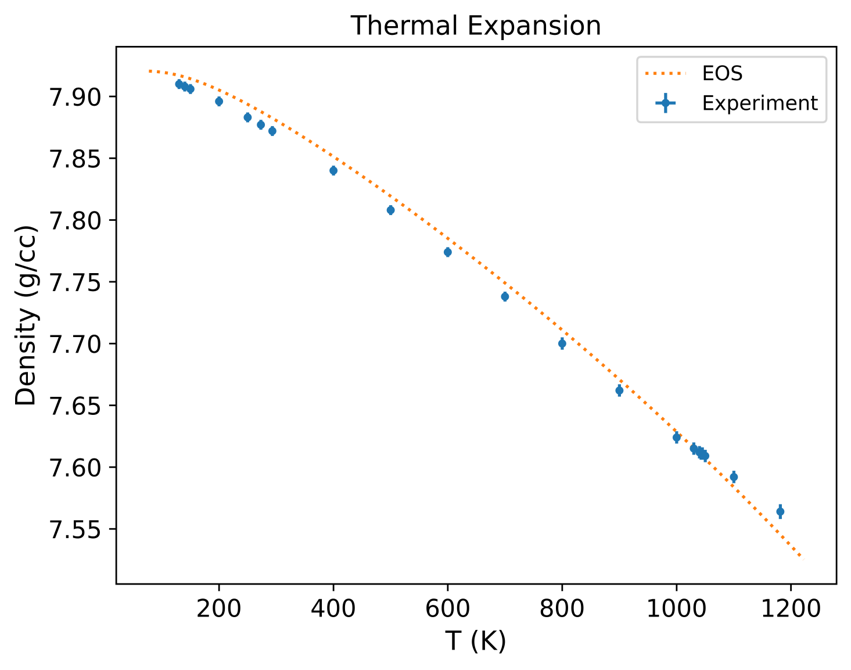}
        \caption*{(d)}
    \end{subfigure}
    
    \caption{Comparison between experimental data and our EOS fits for the $\alpha$ (BCC) phase: (a) Diamond Anvil Cell (DAC) pressure–density data~\cite{Dewaele2008}, (b) entropy–temperature relation from heat capacity measurements~\cite{NIST_JANAF}, (c) principal Hugoniot pressure–density data~\cite{Barker1974,Brown2000}, and (d) thermal expansion from 1-bar measurements~\cite{Abdullaev_2020, Assael2006}.}
    \label{fig:alpha}
\end{figure}

\begin{figure}[H]
    \centering
    \begin{subfigure}[t]{0.48\textwidth}
        \includegraphics[width=\textwidth]{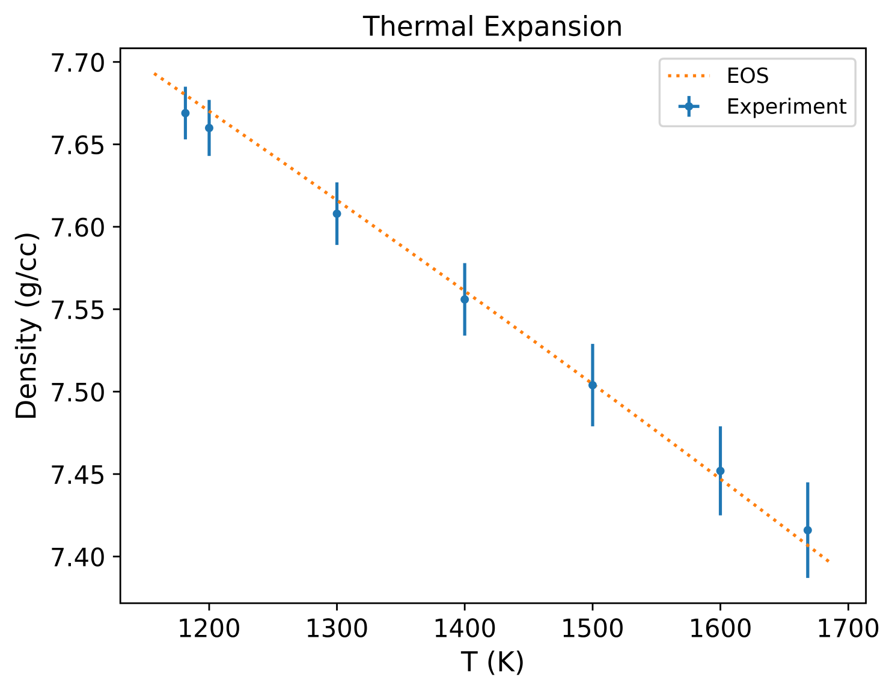}
        \caption*{(a)}
    \end{subfigure}
    \hfill
    \begin{subfigure}[t]{0.48\textwidth}
        \includegraphics[width=\textwidth]{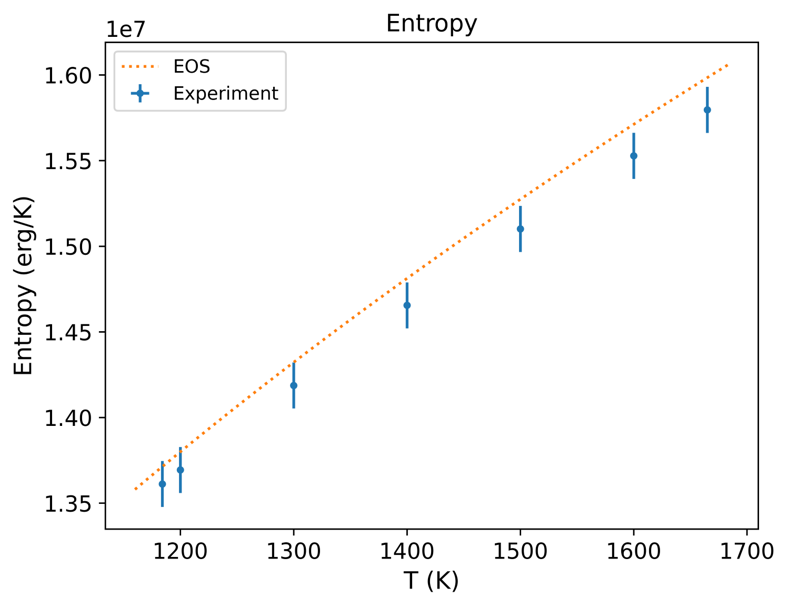}
        \caption*{(b)}
    \end{subfigure}
    
    \vspace{1em}
    
    \begin{subfigure}[t]{0.48\textwidth}
        \includegraphics[width=\textwidth]{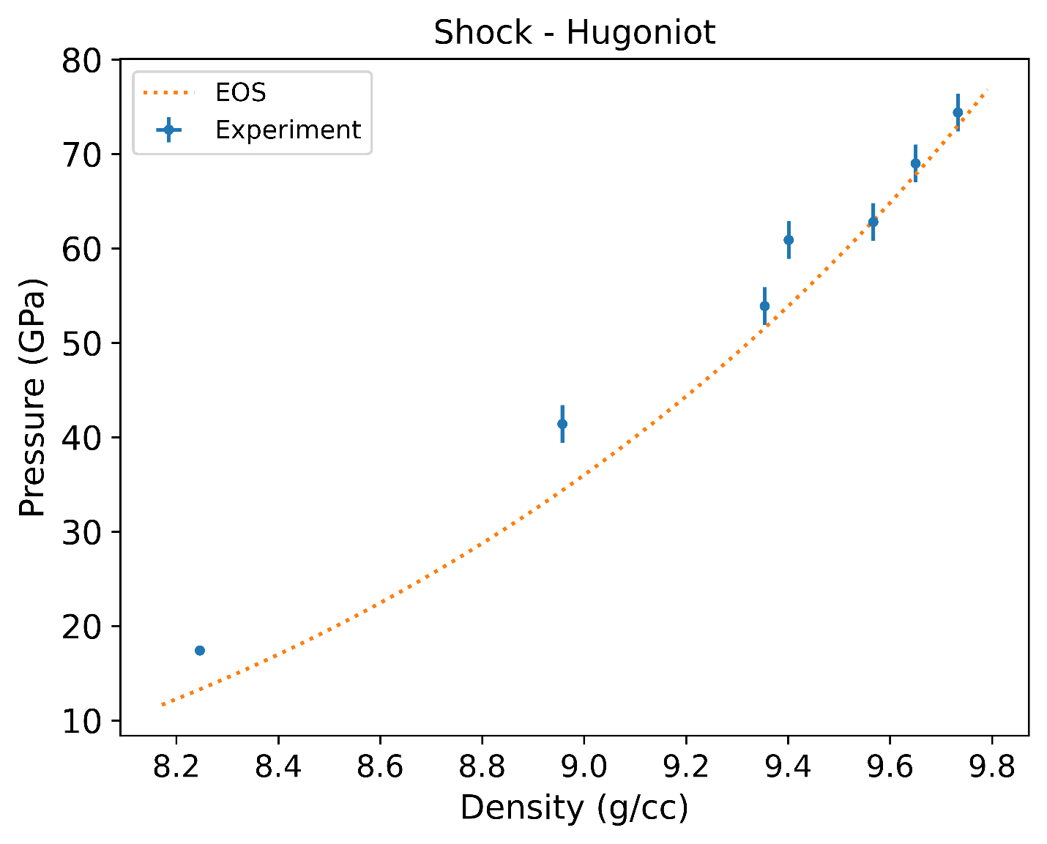}
        \caption*{(c)}
    \end{subfigure}
    \hfill

    \caption{Comparison between experimental data and our EOS fits for the $\gamma$ (FCC) phase: (a) thermal expansion from 1-bar measurements~\cite{Abdullaev_2020, Assael2006}, (b) entropy–temperature relation from heat capacity measurements~\cite{NIST_JANAF}, and (c) heated Hugoniot pressure–density data~\cite{Chen1997}.    }
    \label{fig:gamma}
\end{figure}

\begin{figure}[H]
    \centering
    \begin{subfigure}[t]{0.48\textwidth}
        \includegraphics[width=\textwidth]{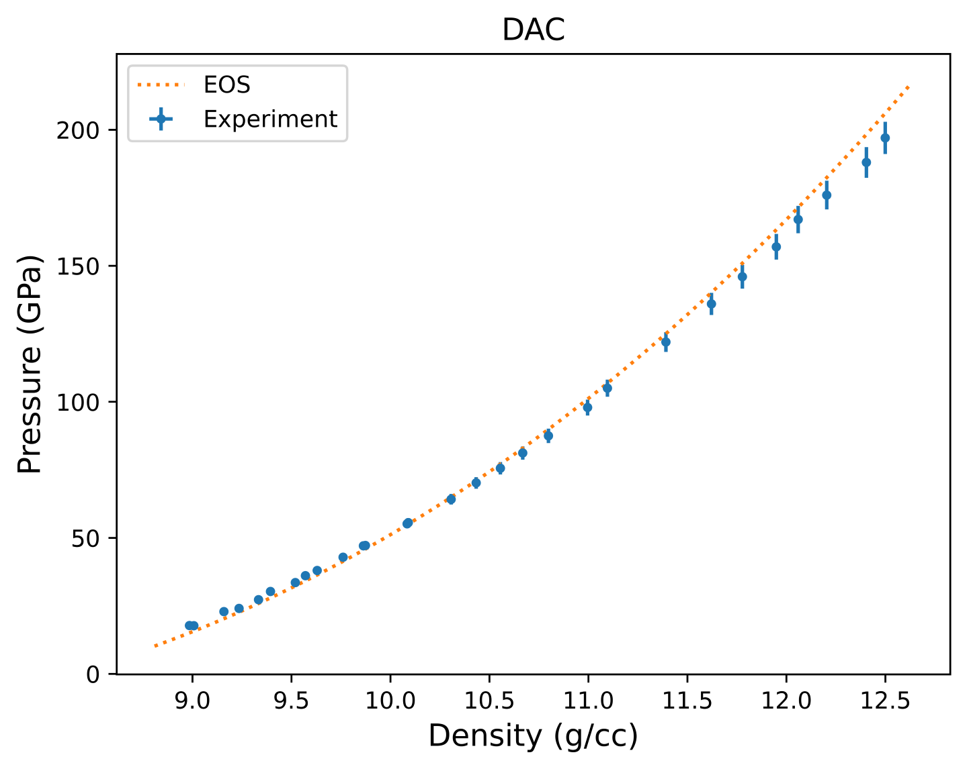}
        \caption*{(a)}
    \end{subfigure}
    \hfill
    \begin{subfigure}[t]{0.48\textwidth}
        \includegraphics[width=\textwidth]{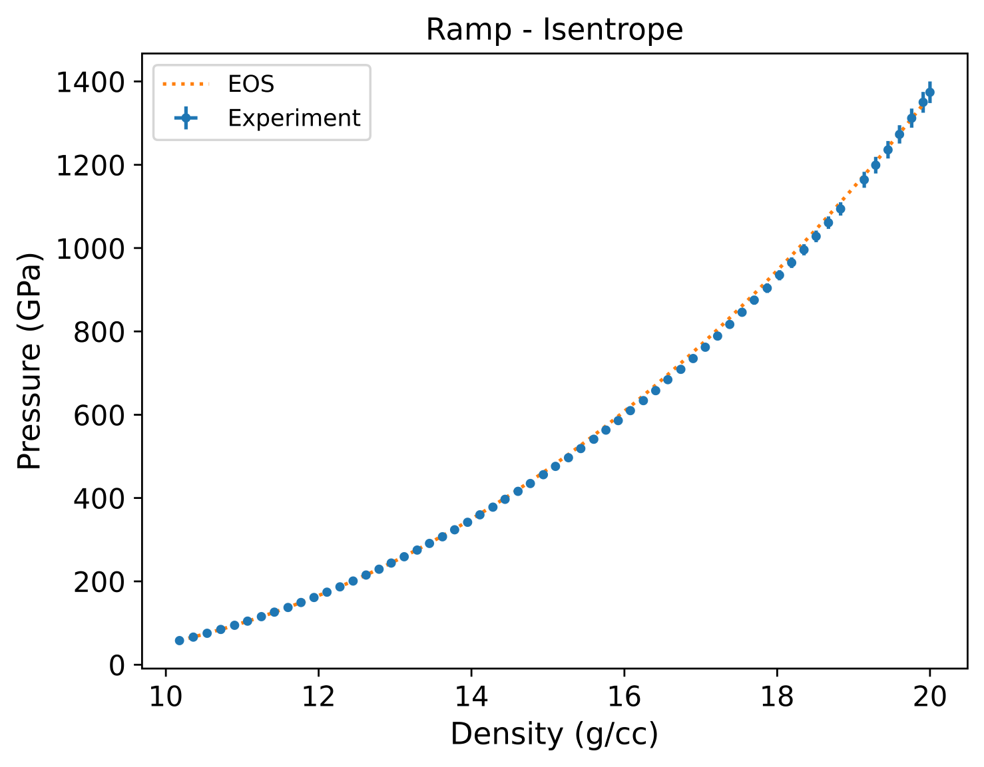}
        \caption*{(b)}
    \end{subfigure}
    
    \vspace{1em}
    
    \begin{subfigure}[t]{0.48\textwidth}
        \includegraphics[width=\textwidth]{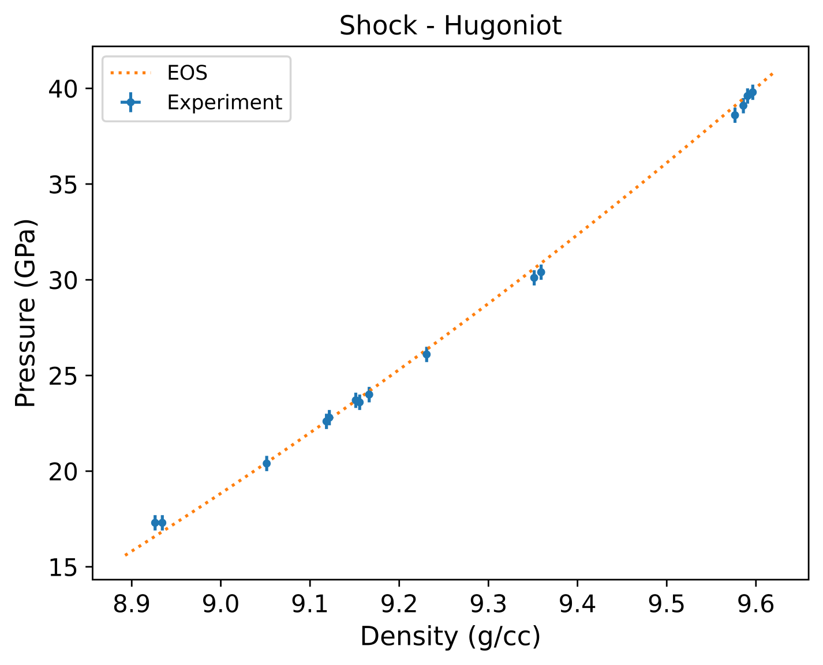}
        \caption*{(c)}
    \end{subfigure}
    \hfill
    \begin{subfigure}[t]{0.48\textwidth}
        \includegraphics[width=\textwidth]{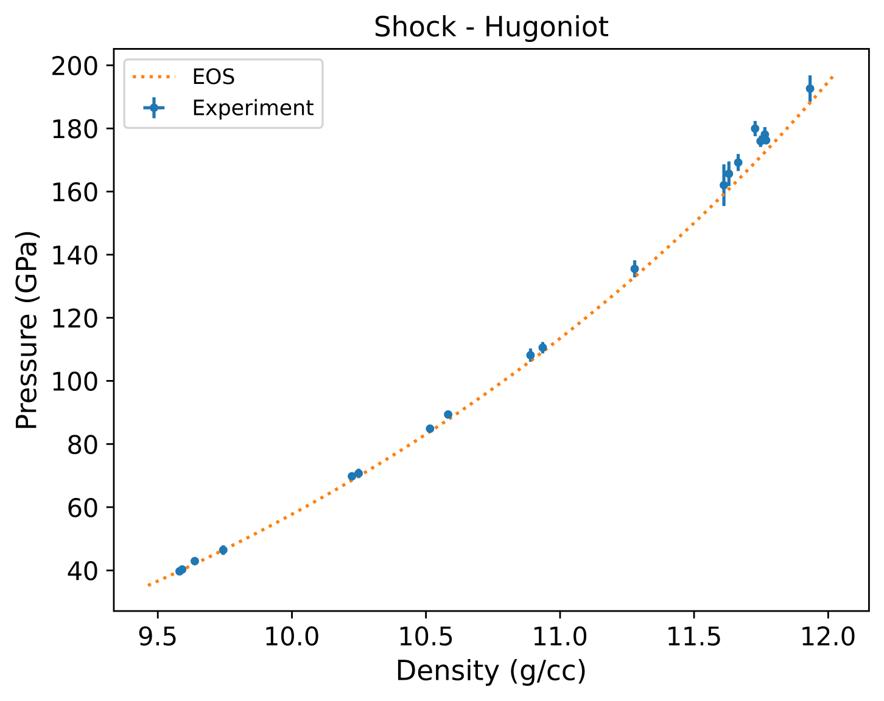}
        \caption*{(d)}
    \end{subfigure}
    
    \caption{Comparison between experimental data and our EOS fits for the $\varepsilon$ (HCP) phase: (a) DAC pressure–density data~\cite{Dewaele2008}, (b) ramp-compression data~\cite{Smith2018}, (c) principal Hugoniot data~\cite{Barker1974,Brown2000}, and (d) high-pressure Hugoniot data from shock measurements.}
    \label{fig:epsilon}
\end{figure}

\begin{figure}[H]
    \centering
    \includegraphics[width=0.8\textwidth]{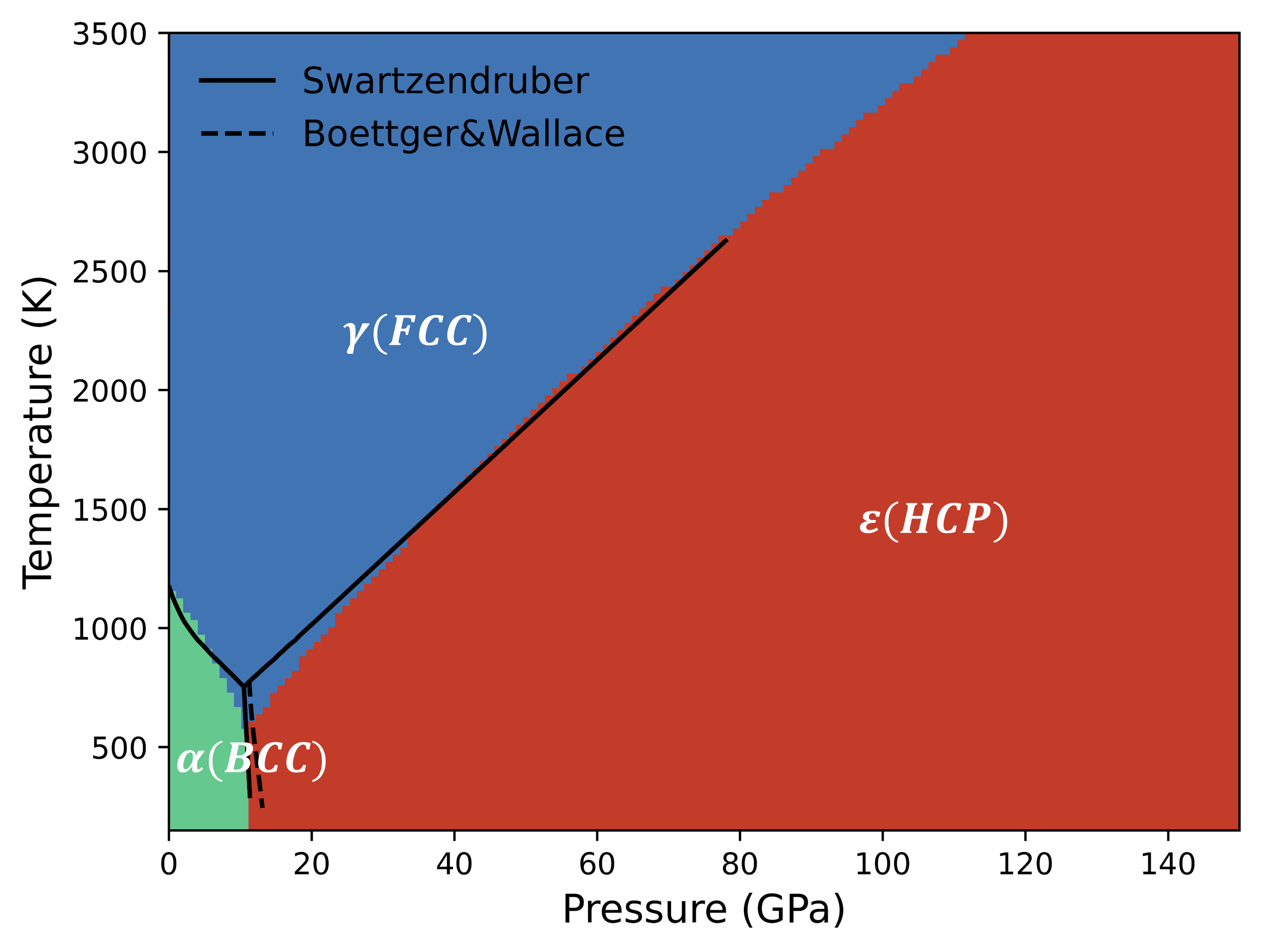}
    \caption{Fe solid phases' phase diagram calculated from the fitted EOS, showing the $\alpha$ (BCC), $\gamma$ (FCC), and $\varepsilon$ (HCP) stability regions, with phase transition lines from Swartzendruber~\cite{Swartzendruber1982} (solid) and Boettger \& Wallace~\cite{Boettgaer1997} (dashed).}
    \label{fig:solid_phases}
\end{figure}

The fitting results are presented in Figs.~\ref{fig:alpha}-~\ref{fig:solid_phases}.
Our EOS accurately reproduces the experimental data of the solid phases. The discrepancy near the solid's triple point is less significant, as the primary purpose of this EOS is to model melting accurately and to enable reliable extrapolation into the liquid regime. 
The discrepancy near the solid's triple point is of lesser importance, as the primary goal of this EOS is to model accurately melting, its region, and allow for reliable extrapolation into the liquid regime.
Furthermore, the parameters closely resemble those found by \cite{Livermore2023}.

\clearpage

\clearpage

\clearpage

\section{Liquid Fitting Parameters}
\label{Appendi:Liquid}

The liquid phase thermodynamic parameters were fitted to the model described in [4]. Table~\ref{tab:liquid} lists the optimized parameters for both FCC-like and HCP-like liquid states.

\begin{table}[H]
    \centering
    \caption{Liquid fitting parameters}
    \begin{ruledtabular}
    \begin{tabular}{cccccccc}
         Phase& $\beta$ & $\alpha_{1}$ & $\alpha_{2}$ & a & b & $\nu$ & $\xi$\\ \hline
         $L_{FCC}$& $55 \pm 5$  & $ 1.10 \pm 0.02$ & $ 0.035 \pm 0.006$ & $ 0.041 \pm 0.001$ & $ 0.101 \pm 0.007$ & $ 4.00 \pm 0.02$ & $ 0.28 \pm 0.015$\\
         $L_{HCP}$ & $ 54.9 \pm 3.5$ & $ 0.97 \pm 0.01$ & $ 0.0005 \pm 0.00002$ & $ 0.371 \pm 0.005$ & $ 0.14 \pm 0.01$ & $ 10.00 \pm 0.01$ & $ 0.07 \pm 0.01$
    \end{tabular}
    \end{ruledtabular}
    
    \label{tab:liquid}
\end{table}

\begin{figure}[H]
    \centering
    \begin{subfigure}[t]{0.48\textwidth}
        \includegraphics[width=\textwidth]{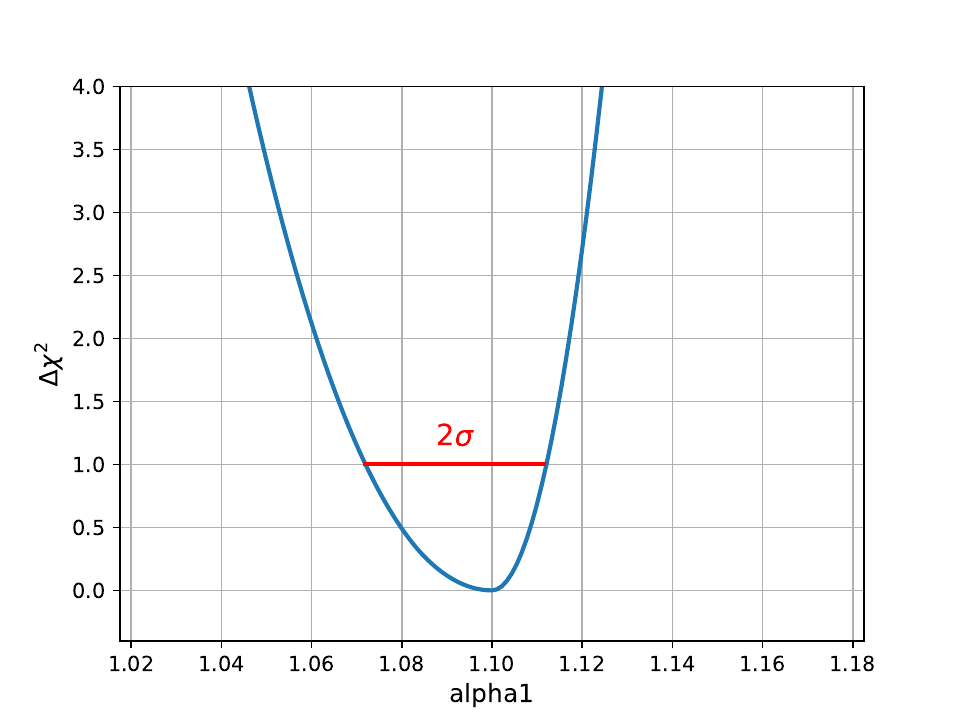}
        \caption*{(a)}
    \end{subfigure}
    \hfill
    \begin{subfigure}[t]{0.48\textwidth}
        \includegraphics[width=\textwidth]{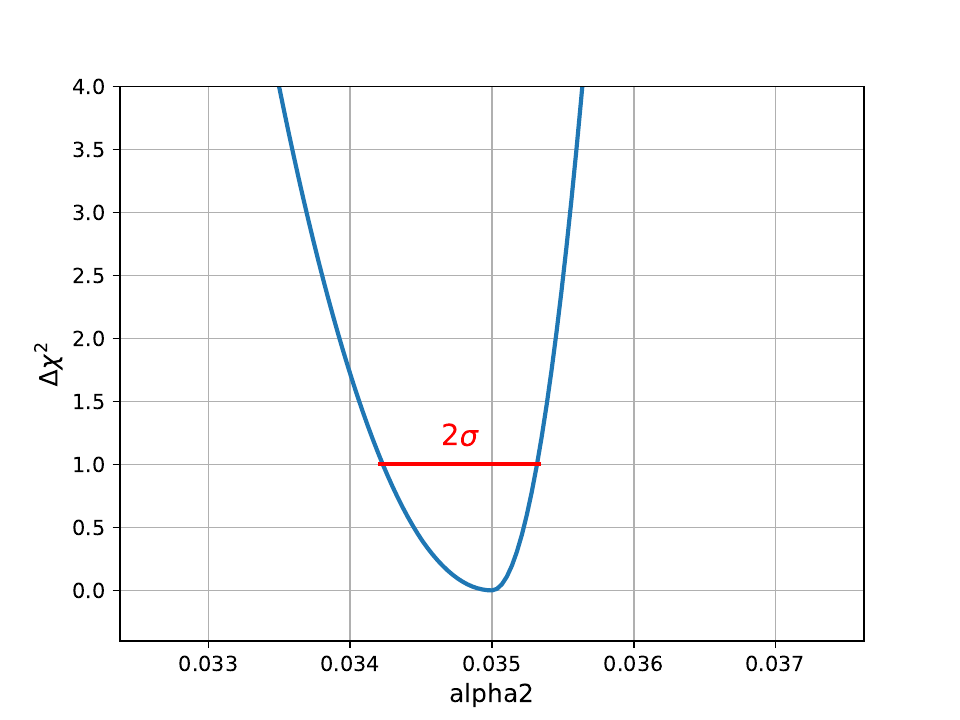}
        \caption*{(b)}
    \end{subfigure}
    
    \vspace{0.5em}
    
    \begin{subfigure}[t]{0.48\textwidth}
        \includegraphics[width=\textwidth]{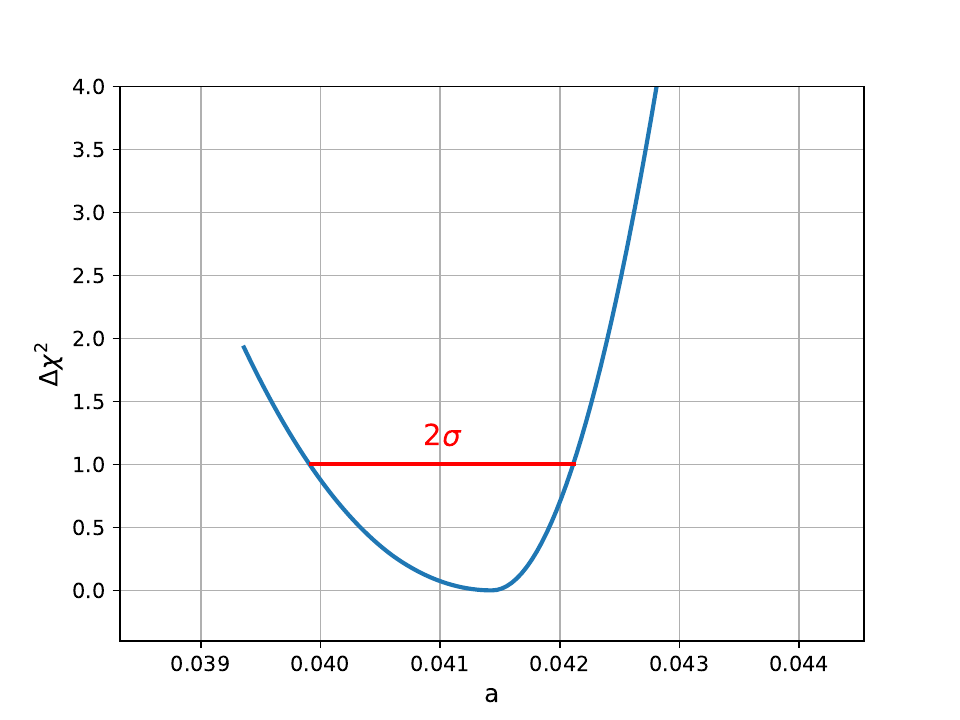}
        \caption*{(c)}
    \end{subfigure}
    \hfill
    \begin{subfigure}[t]{0.48\textwidth}
        \includegraphics[width=\textwidth]{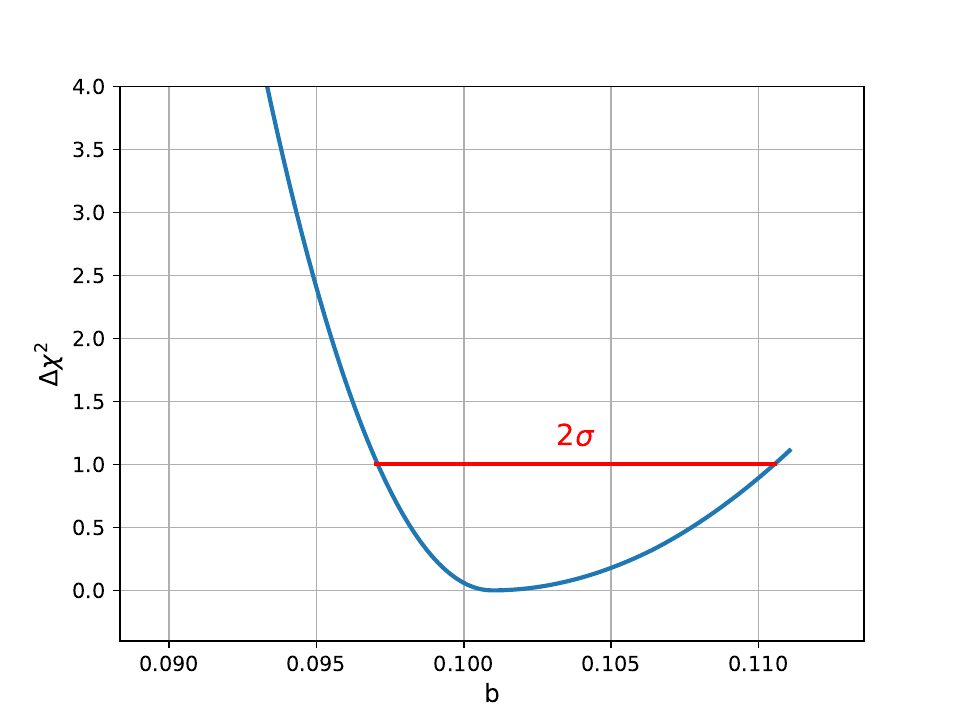}
        \caption*{(d)}
    \end{subfigure}
    
    \caption{Illustration of the uncertainty evaluation procedure. Shown are $\chi^{2}$ variations with respect to four $L_{FCC}$ parameters ($\alpha_1, \alpha_2, a, b$); the same analysis was applied to all liquid-phase parameters. Uncertainties were estimated by identifying the parameter shift corresponding to an increase of $\chi^{2}$ by one unit, referred to here as the $2\sigma$.}
    \label{fig:parameters_error}
\end{figure}

The errors presented here are estimates based on fitting the parameters of the free energies regarding the solid-liquid transition. These error estimates are partial as they do not employ a Hessian covering the complete phase diagram; developing such an estimate is beyond the scope of the current work.  
The error estimates were calculated by varying one parameter at a time and identifying the shift required to increase $\chi^{2}$ by one unit as illustrated in Fig.~\ref{fig:parameters_error}.
This procedure provides only a sensitivity estimate as the error distribution is not Gaussian, and parameter correlations are neglected. 
Moreover, the evaluation is confined to the vicinity of the identified minimum.
Consequently, the quoted $2\sigma$ ranges should be considered indicative rather than strict confidence intervals.

\section{Data Availability}

All data used in this manuscript are provided in \cite{dataset}.

\end{document}